\documentclass{article}

\usepackage{arxiv}
\usepackage[utf8]{inputenc} 
\usepackage[T1]{fontenc}    
\usepackage{hyperref}       
\usepackage{url}            
\usepackage{booktabs}       
\usepackage{amsfonts}       
\usepackage{nicefrac}       
\usepackage{microtype}      
\usepackage{lipsum}
\usepackage{graphicx}
\usepackage{subcaption}
\usepackage{caption}
\usepackage{longtable}
\usepackage{siunitx}
\usepackage{pgfplotstable}
\pgfplotsset{compat=1.18}
\usepackage{multirow}
\usepackage{capt-of}
\usepackage{amsmath, amssymb}  
\usepackage{bm}

\title{A Hybrid Quantum–AI Framework for Protein Structure Prediction on NISQ Devices}

\author{
Yuqi Zhang\thanks{First author; primary contributor.} \\
Kent State University \\
Kent, OH, USA \\
Cleveland Clinic \\
Cleveland, OH, USA \\
\And
Yuxin Yang \\
Cleveland Clinic \\
Cleveland, OH, USA \\
\And
Feixiong Cheng \\
Cleveland Clinic \\
Cleveland, OH, USA \\
\AND
Cheng-Chang Lu \\
Qradle Inc \\
Kent, OH, USA \\
\And
Nima Saeidi \\
Massachusetts General Hospital \\
Boston, MA, USA \\
\And
Samuel L. Volchenboum \\
University of Chicago \\
Chicago, IL, USA \\
\And
Junhan Zhao \\
University of Chicago \\
Chicago, IL, USA \\
\And
Siwei Chen \\
University of Chicago \\
Chicago, IL, USA \\
\And
Weiwen Jiang \\
George Mason University \\
Fairfax, VA, USA \\
\And
Qiang Guan\thanks{Corresponding author.} \\
Kent State University \\
Kent, OH, USA \\
Miami University \\
Oxford, OH, USA \\
\texttt{qguan@kent.edu}
}

\begin{document}

\maketitle

\begin{abstract}
Variational quantum algorithms provide a direct, physics-based approach to protein structure prediction, but their accuracy is limited by the coarse resolution of the energy landscapes generated on current noisy devices. We propose a hybrid framework that combines quantum computation with deep learning, formulating structure prediction as a problem of energy fusion. Candidate conformations are obtained through the Variational Quantum Eigensolver (VQE) executed on IBM’s 127 qubit superconducting processor, which defines a global yet low resolution quantum energy surface. To refine these basins, secondary structure probabilities and dihedral angle distributions predicted by the NSP3 neural network are incorporated as statistical potentials. These additional terms sharpen the valleys of the quantum landscape, resulting in a fused energy function that enhances effective resolution and better distinguishes native like structures. Evaluation on 375 conformations from 75 protein fragments shows consistent improvements over AlphaFold3, ColabFold, and quantum only predictions, achieving a mean RMSD of 4.9~\AA\ with statistical significance ($p<0.001$). The findings demonstrate that energy fusion offers a systematic method for combining data driven models with quantum algorithms, improving the practical applicability of near term quantum computing to molecular and structural biology.
\end{abstract}

\section*{Introduction}

Protein structure prediction is a central challenge in computational biology, with profound implications for understanding biological function and accelerating therapeutic discovery~\cite{onuchic2004theory}. Accurately predicting a protein’s three-dimensional structure from its amino acid sequence provides essential insights into molecular interactions, enzymatic mechanisms, and drug design. In recent years, classical deep-learning methods have achieved remarkable success in this domain, with models capable of delivering highly accurate predictions across large-scale benchmarks~\cite{mirdita2022colabfold}. However, these approaches are inherently data-driven, relying heavily on extensive sequence alignments and training datasets, and they often lack explicit incorporation of physical principles. This limits their interpretability and generalizability, particularly in scenarios that fall outside the training distribution. For example, when modeling specific binding sites or requiring high-resolution predictions for short peptide fragments, deep-learning models often exhibit structural distortions~\cite{scardino2023good}.

Quantum computing offers a fundamentally different paradigm by directly encoding amino acid interactions, chirality constraints, and chemical exclusion principles. Variational quantum algorithms, particularly the Variational Quantum Eigensolver (VQE), have emerged as strong candidates for near-term quantum hardware~\cite{fedorov2022vqe}. VQE optimizes a parameterized quantum circuit to approximate the ground-state energy of a molecular Hamiltonian, thereby providing a physics-based pathway to explore protein folding~\cite{robert2021resource}. Applying VQE to protein fragments opens the possibility of structure prediction from first principles. Nevertheless, current quantum predictions face several challenges: they are vulnerable to hardware noise, they lack biological priors, and due to both modeling granularity and limited qubit resources, existing quantum algorithms often fail to accurately reproduce secondary-structure motifs or backbone dihedral angle distributions~\cite{doga2024perspective}.

To address these limitations, we incorporate biological information derived from data analysis into the quantum modeling process. Classical deep learning methods, although not without constraints, are highly effective at recognizing recurring patterns in local protein geometry, such as secondary structure motifs and characteristic torsion angle distributions, which capture empirical regularities drawn from extensive structural datasets. These correlations are generally inaccessible to quantum formulations that operate under restricted qubit resources and simplified Hamiltonian representations. Building on this complementarity, we introduce a unified framework that combines the physics based modeling capability of the Variational Quantum Eigensolver (VQE) with biological priors obtained from neural network predictions. In this formulation, the quantum computation generates candidate conformations subject to physical constraints, while the neural priors serve as structural references that refine and reorder the quantum results toward biologically consistent protein folds.

From the perspective of conformational energy landscapes, this design can be viewed as the construction of a fused energy function that modifies the quantum energy surface. The quantum energy landscape offers a reliable but coarse map of low energy basins, whereas the inclusion of biological priors obtained from learning models introduces finer gradients that sharpen these basins and effectively enhance their resolution. The resulting fused landscape preserves the global consistency of quantum mechanics while gaining the local discriminative ability of data derived potentials, allowing the identification of native like structures that would otherwise remain indistinguishable within broad quantum minima~\cite{frauenfelder1991energy}.

This design moves beyond approaches that rely solely on data-driven inference or purely on physical simulation, demonstrating how domain knowledge from deep learning can enhance the reliability of noisy quantum computations. By combining the complementary strengths of quantum and classical paradigms, our method provides a feasible pathway for tackling complex molecular modeling tasks on near-term quantum hardware.

The main contributions of this work are as follows:
\begin{itemize}
    \item We design a hybrid framework that integrates VQE with predicted priors of deep-learning on secondary structures and dihedral angles.
    \item We introduce an energy reweighting mechanism that refines quantum predictions, yielding protein conformations that are biologically meaningful.
    \item We validate our method on 75 protein fragments generated on real quantum processors, demonstrating consistent RMSD improvements over both classical and quantum baselines.
    \item We highlight a general strategy for combining deep-learning priors with quantum algorithms, providing a practical pathway toward near-term quantum applications in computational biology.
\end{itemize}

\section*{Background}

\subsection*{Protein Structure Prediction}

Accurately predicting protein structures from amino acid sequences is one of the central challenges of modern science.~\cite{whisstock2003prediction} Proteins underpin nearly all biological processes, and their three-dimensional conformations determine their functions, interactions, and dynamics~\cite{dorn2014three}. Reliable structure prediction therefore has broad implications, spanning fundamental life sciences, medicine, and therapeutic development. In the pharmaceutical domain, structural insights guide the rational design of small molecules and biologics to selectively modulate biological pathways. In life sciences, structural knowledge is essential for understanding enzymatic mechanisms, cellular signaling, and disease-associated mutations~\cite{copeland2023enzymes}. Beyond biology and medicine, protein structure prediction has also emerged as a benchmark problem in computer science and artificial intelligence, driving advances in deep learning architectures, large-scale data processing, and high-performance computing. Progress in this field thus simultaneously deepens our understanding of biology and pushes the frontiers of computational methodology~\cite{xu2007computational}.

In recent years, several methodological paradigms have been developed for protein structure prediction, each with distinct principles and limitations. Template-based modeling (TBM) relies on the structural similarity between the query sequence and proteins of known structure. While TBM can provide accurate predictions when close homologs exist in the Protein Data Bank (PDB), it struggles for novel folds or low-homology sequences~\cite{szilagyi2014template}. Ab initio or physics-based methods, on the other hand, attempt to identify the native conformation by directly minimizing an energy function derived from molecular force fields~\cite{hardin2002ab}. These approaches are grounded in physical principles and can in principle handle novel proteins, but they are computationally expensive and limited by the accuracy of current force fields.

The most substantial progress has been achieved through deep learning methods, represented by AlphaFold2 and AlphaFold3 as well as related frameworks~\cite{mirdita2022colabfold}. These approaches make use of large multiple sequence alignments, coevolutionary information, and attention based neural architectures to predict inter residue distances, orientations, and complete three dimensional coordinates. Their main advantages are the high accuracy obtained across diverse protein families and the ability to scale to whole proteomes~\cite{yang2024deep}. Nevertheless, their dependence on extensive training data and large sequence alignments can reduce effectiveness for orphan proteins with few homologs, and their predicted structures, although often geometrically consistent, may not always represent true energetic stability or ligand binding capability.

From a theoretical perspective, protein structure prediction can also be interpreted through the lens of the \textbf{energy landscape}. An amino acid sequence defines a high-dimensional potential energy surface, where each point corresponds to a possible conformation~\cite{neelamraju2020protein}. The topology of this landscape governs the folding pathways and final outcomes: the global minimum corresponds to the native folded state, whereas local minima capture metastable or misfolded states. Traditional physics-based approaches attempt to search this landscape explicitly using molecular mechanics or quantum models, but are hindered by computational complexity. Deep-learning methods, in contrast, learn statistical regularities of the landscape from large datasets, transforming the search into a pattern recognition problem within a constrained space. Hybrid approaches, such as the one proposed in this study, aim to combine the physical fidelity of quantum models with the statistical priors of deep learning to more efficiently and accurately approximate the true energetic minima.

The performance of protein structure prediction methods is commonly assessed using quantitative metrics. The most widely adopted is the \emph{root-mean-square deviation (RMSD)} of atomic positions, typically computed over backbone or C$\alpha$ atoms relative to experimental structures. RMSD provides a straightforward measure of structural similarity: lower values indicate closer alignment to the native structure. For instance, an RMSD below 5\,\AA\ is often considered near-atomic accuracy, while values above 8--10\,\AA\ indicate substantial deviation~\cite{carugo2003root}. Other complementary metrics include the Template Modeling score (TM-score), which normalizes for protein length and better distinguishes correct folds from random similarities. Among these, RMSD remains the most interpretable and intuitive, directly reflecting the geometric fidelity of predicted models to experimental ground truth.

\subsection*{Quantum Computing in Protein Structure Prediction}

Quantum computing offers a fundamentally different paradigm for addressing the protein-structure problem by formulating it as a ground-state (or low-lying eigenstate) search on a rugged, high-dimensional \emph{energy landscape}~\cite{marchetti2022quantum}. Let $\mathcal{X}$ denote the space of protein conformations (e.g., backbone torsion assignments or coarse-grained lattice states), and let $E(x)$ be the classical potential energy associated with $x\in\mathcal{X}$. A quantum algorithm encodes this landscape into a Hermitian operator $\hat H$ acting on a Hilbert space spanned by computational basis states $\{\lvert x\rangle\}$ that correspond to discrete conformations, so that the expectation $\langle x\vert \hat H \vert x\rangle\approx E(x)$. The task becomes preparing a parameterized quantum state $\lvert \psi(\boldsymbol\theta)\rangle$ and minimizing
\[
\min_{\boldsymbol\theta}\; \langle \psi(\boldsymbol\theta)\vert \hat H \vert \psi(\boldsymbol\theta)\rangle,
\]
so that the state’s support concentrates on low-energy basins.

From the perspective of the energy-landscape picture, quantum computing exploits several uniquely quantum features. Superposition allows $\lvert \psi\rangle$ to represent an exponentially large ensemble of conformations at once, distributing amplitudes across many basins of the landscape. Entanglement captures high-order correlations among residues (e.g., long-range contacts, cooperative packing) that are difficult to encode with low-order classical features. In the idealized picture, interference steers amplitudes toward low-energy valleys while suppressing high-energy regions. In practice, however, no measurement can “read out” all configurations simultaneously; algorithmic mechanisms must therefore \emph{reweight} amplitudes so that low-energy structures are sampled with high probability~\cite{pal2024quantum}.

These ideas are instantiated through variational algorithms such as VQE and QAOA~\cite{choi2019tutorial}. VQE prepares $\lvert \psi(\boldsymbol\theta)\rangle=U(\boldsymbol\theta)\lvert 0\rangle$ with an ansatz of limited depth and estimates $E(\boldsymbol\theta)=\sum_j w_j \langle \psi\vert P_j\vert \psi\rangle$ by decomposing $\hat H=\sum_j w_j P_j$ into Pauli strings and performing repeated measurements~\cite{tilly2022variational}. A classical optimizer (gradient-free or parameter-shift gradients) updates $\boldsymbol\theta$ to descend the energy landscape. QAOA, by contrast, alternates cost and mixer unitaries, $U(\boldsymbol\gamma,\boldsymbol\beta)=\prod_{\ell=1}^{p}\exp(-i\beta_\ell \hat M)\exp(-i\gamma_\ell \hat H)$, where increasing depth $p$ enriches the landscape navigation capacity~\cite{zhou2020quantum}. Both families rely on interference to bias the state toward low-energy basins rather than enumerating the landscape exhaustively.

A key design question is how to construct Hamiltonians and encodings. Different approaches offer distinct trade-offs between physics fidelity and qubit requirements:  
(i) backbone torsion encodings discretize $(\phi,\psi)$ into $K_\phi, K_\psi$ bins per residue and encode indices in binary, leading to a linear scaling in qubits with residue count;  
(ii) coarse-grained contact or Ising-type models encode residue–residue interactions as 2-local couplings, yielding shallow cost operators but sacrificing atomic detail;  
(iii) electronic-structure–inspired encodings treat small active sites with fermionic mappings, providing high accuracy but at the cost of severely limited system size. Because protein energetics involves long-range, many-body couplings, compiling to hardware with limited connectivity introduces SWAP overhead and depth growth, constraining the feasible model class.

It is important to clarify why quantum “instant exploration” is not black magic. Superposition lets a circuit represent exponentially many conformations simultaneously, and entanglement encodes their correlations; yet extracting information requires measurements whose variance scales as $\mathcal{O}(1/\sqrt{S})$ with $S$ shots~\cite{ding2024exploring}. Any quantum advantage thus hinges on preparing states whose amplitudes are already concentrated on low-energy regions. Amplitude amplification, interference, and carefully chosen mixers/costs provide this bias—but they do not literally “search all states at once” in a readout sense. Quantum annealing and adiabatic paradigms add the intuition of tunneling through tall, narrow barriers, potentially escaping traps that impede classical thermal search; nevertheless, useful speedups depend on barrier structure and instance properties, and require mappings that preserve landscape topology.

When implemented on real devices, additional challenges arise for trainability and robustness. Barren plateaus can cause gradient magnitudes to decay exponentially with system size or depth, especially for unstructured ansätze; structured or problem-inspired ansätze and locality-aware cost partitions help alleviate this. Shot noise and the large number of Pauli terms in $\hat H$ make energy estimation costly, motivating commuting-group measurement and importance sampling. Hardware noise from decoherence, SPAM, and two-qubit gate errors further bias estimates and limit depth, while error mitigation methods (e.g., zero-noise extrapolation, symmetry verification, virtual distillation) improve fidelity at additional cost. Finally, limited qubit connectivity inflates circuit depth via SWAP networks, degrading trainability on NISQ hardware~\cite{liu2025quclear,luo2025digital}.

These constraints impose granularity limits on today’s protein encodings. Angle discretization reduces qubit requirements but smooths away subtle energy barriers and secondary-structure preferences. Backbone-only or lattice models omit side-chain packing and hydrogen-bond geometry, obscuring helices and strands. Finite-shot estimators and device noise set a floor on resolvable energy differences, making it hard to distinguish near-degenerate conformers. Constraint handling also adds $k$-local interactions, which increase depth or ancilla overhead after reduction to 2-local form. Collectively, these factors explain why present-day VQE and QAOA predictions are often coarse-grained: they locate plausible low-energy \emph{regions} of the landscape but lack explicit helices, sheets, and accurate $(\phi,\psi)$ detail.

Despite these limitations, recent years have seen significant progress in quantum computing for protein structure prediction. A variety of algorithms—ranging from VQE~\cite{malone2022towards} and QAOA~\cite{chen2025hqqubo} to tensor-network–inspired ansätze and quantum Monte Carlo~\cite{austin2012quantum} methods—have been explored to model conformational energy landscapes and identify low-energy states. These approaches demonstrate that quantum processors can encode protein structural features directly into Hamiltonians and perform energy minimization tasks that are otherwise challenging for classical heuristics. More importantly, proof-of-principle studies on current NISQ devices have already shown that quantum algorithms are capable of delivering predictive accuracy surpassing that of state-of-the-art deep-learning models in certain regimes~\cite{zhang2025prediction,zhang2025qdockbank}. For example, quantum–classical hybrid workflows executed on real hardware have achieved lower RMSD values and more consistent docking poses than leading AI-based predictors, underscoring the unique advantage of physics-grounded quantum approaches. Collectively, these results establish that even with the hardware limitations of today’s processors, quantum computing has matured to the point of providing competitive, and in specific scenarios superior, predictive power for protein structure modeling.

These challenges call for approaches that combine physical modeling with new computational resources. Quantum computation provides a natural way to explore the vast conformational landscape of proteins by exploiting effects such as superposition and entanglement, which allow the system to sample multiple structural configurations simultaneously. This capability aligns well with the energy-landscape perspective of protein folding, where low-energy conformations correspond to stable structures. Yet, the current generation of quantum hardware remains constrained by limited qubit numbers, circuit noise, and shallow circuit depth. As a result, only coarse-grained encodings can be implemented, and the attainable energy resolution is restricted. To overcome these barriers, hybrid pipelines combine quantum-generated candidates that capture the underlying low-energy physics with deep-learning priors that provide structural knowledge such as secondary-structure motifs, dihedral angles, and solvent exposure. In our framework, candidate conformations are first generated using VQE on real quantum processors to capture fundamental physical interactions. These conformations are then refined by incorporating deep-learning–predicted secondary-structure probabilities and backbone dihedral angle distributions as weighted energy terms. This integration enhances the structural completeness of quantum predictions while maintaining consistency with fundamental physical constraints.

By combining quantum algorithms grounded in physical theory with artificial intelligence models that capture statistical patterns of protein structure, our framework brings together the strengths of both domains. The artificial intelligence component provides structural cues, such as common tendencies in secondary structure and backbone angles, which help the quantum procedure focus its search on realistic regions of the energy landscape. Through this cooperation, the method overcomes the separate limitations of classical learning and quantum simulation, offering a practical way forward for the use of quantum computing in structural biology. To our knowledge, this is the first demonstration of protein structure prediction carried out on real quantum hardware, producing physically consistent and complete conformations. More broadly, it offers a template for combining artificial intelligence and quantum computation in molecular modeling.

\section*{Methods}

\subsection*{Framework Overview}

\begin{figure}[t]
    \centering
    \includegraphics[width=0.95\linewidth, trim=0 45pt 0 0, clip]{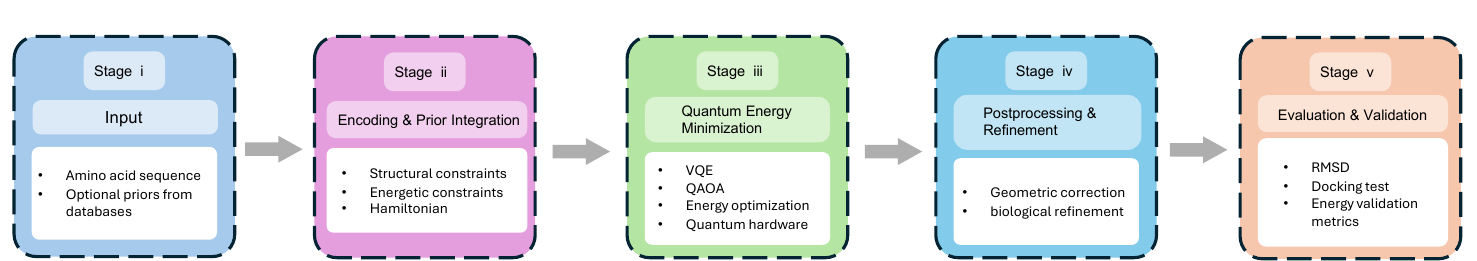}
    \caption{Overview of the five-stage framework for quantum protein structure prediction. 
    The workflow proceeds from sequence input through encoding, quantum energy minimization, refinement, and evaluation. 
    Stages (ii) and (iv) enable hybrid operation by incorporating biologically informed priors derived from neural models. 
    This systematic design provides a reproducible engineering structure that unifies pure quantum prediction and hybrid quantum and classical strategies.}
    \label{fig:five_stage_framework}
\end{figure}

In quantum engineering, protein structure prediction follows a fixed methodological workflow, since the essence of the task is to identify the lowest-energy conformation. We abstract this process into a unified five-stage framework, which applies equally to pure quantum prediction and naturally extends to hybrid quantum--classical strategies. The central idea is to start from the sequence input, integrate prior knowledge into the encoding, perform energy minimization on a quantum processor, refine the resulting structures, and finally validate them with multi-level evaluation metrics. The five stages are: (i) input of the amino acid sequence and optional priors; (ii) prior integration and encoding into a quantum-representable Hamiltonian; (iii) quantum energy minimization using algorithms such as VQE or QAOA; (iv) postprocessing and refinement to ensure geometric and biological validity; and (v) evaluation and validation through structural, energetic, and functional benchmarks. In summary, this five-stage framework provides a repeatable engineering methodology for protein structure prediction with quantum computing, while offering a unified reference point for both pure quantum and hybrid strategies. Figure~\ref{fig:five_stage_framework} summarizes the overall workflow of our proposed five-stage framework for quantum protein structure prediction. The process begins with the sequence input, where amino acid sequences and optional biological priors are prepared as model inputs. In the encoding stage, these priors are integrated into a quantum Hamiltonian that captures the relevant physical and geometric constraints. Quantum energy minimization is then performed using variational algorithms such as VQE or QAOA to identify candidate low energy conformations. The refinement stage applies geometric and biological corrections to improve structural plausibility, followed by the evaluation and validation stage, where the resulting conformations are assessed through structural, energetic, and functional benchmarks. This systematic workflow provides a reproducible engineering structure that applies equally to pure quantum prediction and hybrid quantum and classical implementations.

Building on this general framework, our work develops a hybrid quantum--deep learning pipeline that integrates candidate conformations generated by the Variational Quantum Eigensolver (VQE) with structural features provided by neural predictors. The goal is to construct a re-ranking mechanism that balances the physical fidelity of quantum predictions with structural constraints learned from large-scale protein datasets.  

\paragraph{Quantum candidate generation.}
Given an amino-acid fragment, we encode the sequence into a Hamiltonian that captures steric exclusion and residue interactions, and solve it using a VQE executed on quantum hardware. The VQE outputs a set of candidate conformations 
$\mathcal{C}=\{c_1,\dots,c_N\}$ with corresponding quantum energies $E_q(c)$. These candidates represent physically plausible C$\alpha$ backbones with low energy under the quantum model, but they lack explicit representations of secondary structures and dihedral angles~\cite{zhang2025prediction}.

\paragraph{Deep-learning feature extraction.}
In parallel, we run the neural predictor NetSurfP~\cite{klausen2019netsurfp,hoie2022netsurfp} on the same sequence. For each residue, the model provides three categories of structural priors: (i) SS3/SS8 secondary-structure probabilities $\mathbf{p}^{\,\mathrm{ss}}_i$, (ii) dihedral angles $(\phi_i,\psi_i)$, and (iii) relative solvent accessibility $w_i$. These priors serve as structural constraints and reference distributions that complement the quantum candidates.

\paragraph{Candidate augmentation.}
For each quantum candidate $c$, we extract its dihedral angles $(\hat\phi_i(c),\hat\psi_i(c))$. If explicit dihedrals are unavailable, ``virtual'' dihedrals are computed from the C$\alpha$ coordinates using four-point torsion formulas. From these angles, candidate SS3/SS8 distributions $\hat{\mathbf{p}}^{\,\mathrm{ss}}_i(c)$ are further generated via Ramachandran kernels. This augmentation ensures that each candidate is annotated with both continuous features (angles) and discrete features (secondary-structure distributions), enabling direct comparison with deep-learning priors~\cite{ramachandran1968conformation}.

\paragraph{Feature fusion.}
We define a fused energy $E_{\mathrm{fuse}}(c)$ that linearly combines three terms: (i) normalized quantum energy $\tilde E_q(c)$, (ii) secondary-structure distribution divergence $\tilde D_{\mathrm{ss}}(c)$, and (iii) dihedral-angle consistency $\tilde D_{\angle}(c)$, with trade-off parameters $(\alpha,\beta,\gamma)$. Angle consistency can be optionally weighted by RSA to emphasize residues more likely to be solvent-exposed~\cite{rohl2004protein}. All terms are min--max normalized across candidates to ensure comparability.

\paragraph{Re-ranking and selection.}
For all candidates, we compute $E_{\mathrm{fuse}}(c)$ and rank them in ascending order. The candidate with the lowest fused energy is selected as the final prediction. This step balances the physical fidelity of quantum outputs with the structural priors from deep learning, resulting in conformations that are both energetically favorable and feature-complete.

\begin{figure}[t]
    \centering
    \includegraphics[width=0.95\linewidth]{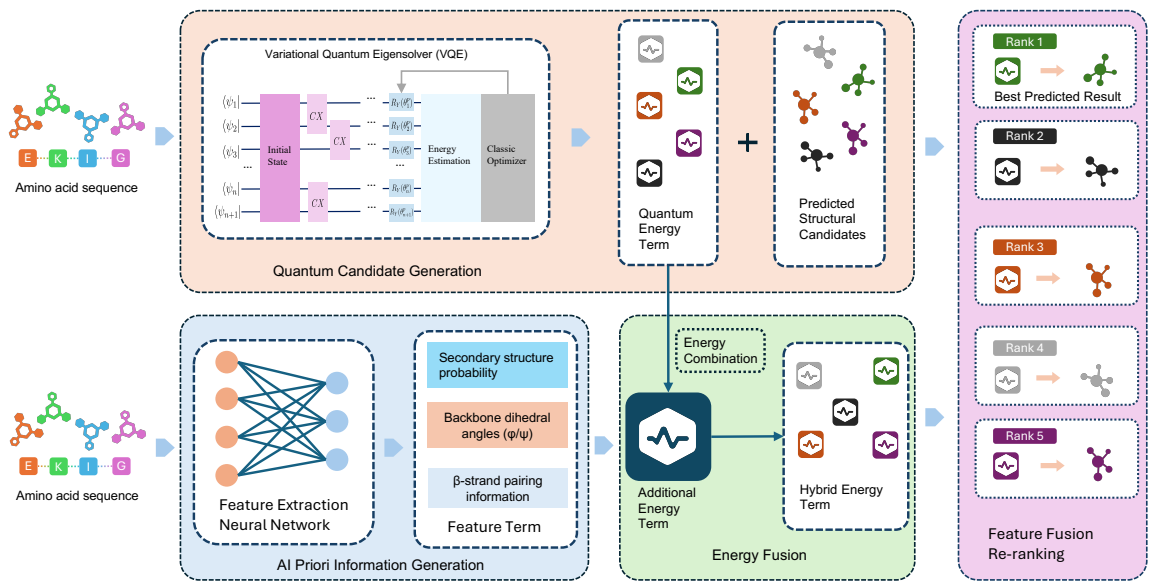}
    \caption{Overview of the proposed hybrid framework. Candidate 
    conformations are generated using VQE on quantum hardware, 
    augmented with structural features from deep learning models, 
    and re-ranked via fused energy terms to produce the final prediction.}
    \label{fig:framework}
\end{figure}

In summary, Our framework achieves a novel integration of quantum computing and artificial intelligence paradigms: the quantum component provides physically grounded candidate conformations, while the deep-learning component supplies structural priors that correct and enrich these candidates. Through the re-ranking mechanism, the two perspectives are unified into a consistent scoring scheme, enabling reliable and robust prediction of protein fragment structures that are otherwise difficult to obtain using either method alone. An overview of the pipeline is illustrated in Figure~\ref{fig:framework}, which depicts the flow from quantum candidate generation to deep-learning feature extraction, feature fusion, and final re-ranking.

\subsection*{Principles of Hybrid Energy Landscapes}

\begin{figure*}[h]
    \centering
    \begin{subfigure}[t]{0.32\textwidth}
        \centering
        \includegraphics[width=\textwidth]{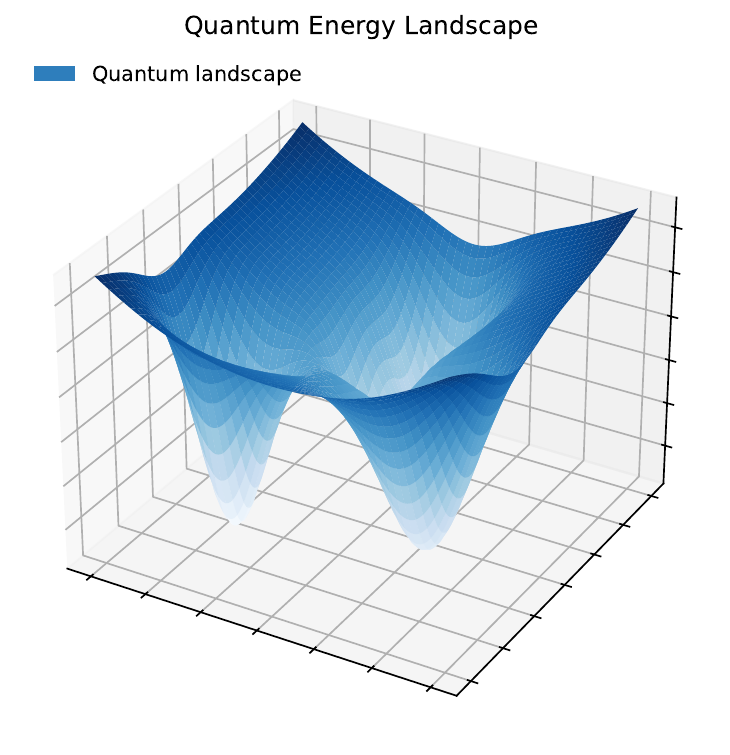}
        \caption{Quantum energy landscape with multiple basins.}
        \label{fig:landscape_quantum}
    \end{subfigure}
    \begin{subfigure}[t]{0.32\textwidth}
        \centering
        \includegraphics[width=\textwidth]{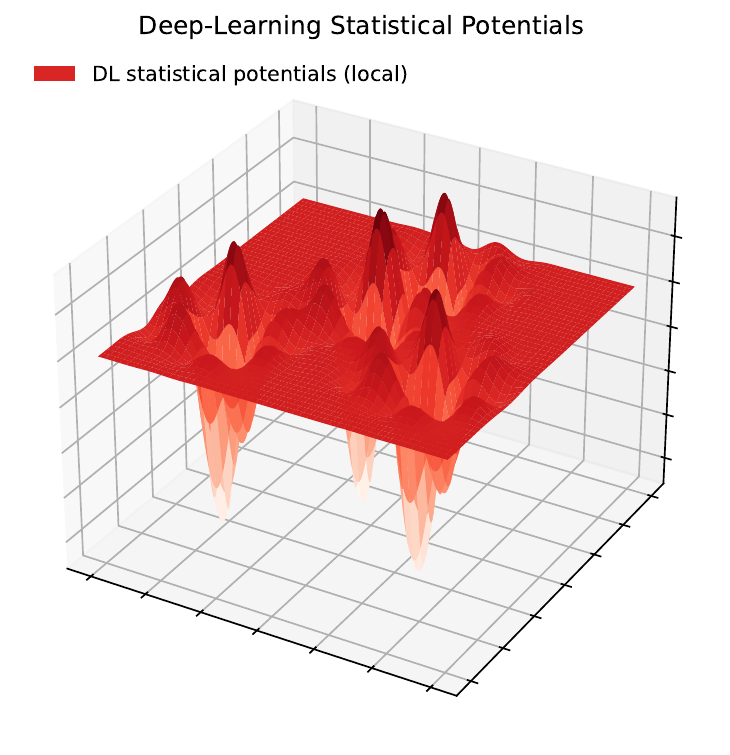}
        \caption{Deep-learning statistical potentials providing local refinements.}
        \label{fig:landscape_dl}
    \end{subfigure}
    \begin{subfigure}[t]{0.32\textwidth}
        \centering
        \includegraphics[width=\textwidth]{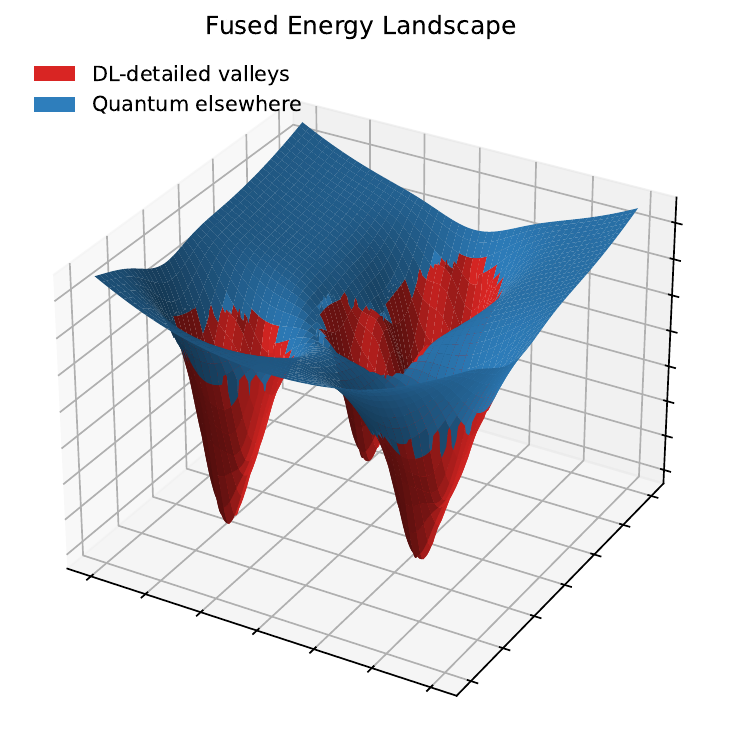}
        \caption{Fused landscape: DL-detailed valleys and quantum elsewhere.}
        \label{fig:landscape_fused}
    \end{subfigure}
    \caption{Comparison of energy landscapes. 
    (a) Quantum-derived energy landscape shows multiple coarse-grained basins. 
    (b) Deep-learning priors contribute fine-grained local refinements in valley regions. 
    (c) The fused energy landscape sharpens valley resolution by combining quantum basins with DL-detailed gradients, enabling more accurate identification of native-like structures.}
    \label{fig:energy_landscapes}
\end{figure*}

The rationale behind our framework can be understood most clearly by
considering protein structure prediction as a problem of navigating
the conformational energy landscape. In the canonical formulation, a
protein sequence defines a high-dimensional potential energy surface,
where each point corresponds to a possible conformation and the
topology of the surface determines the folding process. The global
minimum corresponds to the native folded state, while local minima
capture alternative metastable conformations. Figure~\ref{fig:energy_landscapes}
illustrates this perspective in terms of three complementary energy
landscapes: quantum-derived, deep-learning–based, and their fused
integration.

\paragraph{Quantum energy landscape.}
Variational quantum algorithms such as VQE provide a principled
first-principles approach to approximating this energy landscape by
encoding steric exclusion, residue interactions, and other relevant
Hamiltonian terms into quantum operators. The resulting landscape
reflects fundamental quantum-mechanical interactions but is inherently
coarse-grained under current hardware and algorithmic limitations.
In practice, quantum-derived energy surfaces capture the broad basins
of attraction—the approximate locations of low-energy conformations—
but cannot resolve fine-grained features such as local torsional
preferences or precise secondary-structure motifs. The quantum
landscape therefore provides a reliable but low-resolution map, as
depicted in Fig.~\ref{fig:energy_landscapes}a.

\paragraph{Deep-learning energy landscape.}
By contrast, deep-learning predictors such as NetSurfP are trained on
large corpora of experimentally determined protein structures. Their
outputs can be interpreted as statistical energy-like potentials:
secondary-structure probabilities, backbone dihedral angle
distributions, and solvent accessibility values implicitly encode the
statistical preferences observed in real proteins. This learned
landscape is high-resolution in terms of local structural features but
does not emerge from first-principles energetics. As a consequence,
deep learning can supply sharp gradients in conformational space,
pinpointing preferred torsional angles and motif arrangements, yet its
global validity is not guaranteed in data-sparse or atypical
environments. Figure~\ref{fig:energy_landscapes}b highlights this
local refinement effect.

\paragraph{Hybrid energy formulation.}
Our framework unifies these two complementary landscapes by defining a
fused energy function
\[
E_{\mathrm{fuse}}(c) = \alpha \,\tilde{E}_q(c) +
  \beta \,\tilde{D}_{\mathrm{ss}}(c) +
  \gamma \,\tilde{D}_{\angle}(c),
\]
where the normalized quantum energy $\tilde{E}_q$ provides the
low-resolution global structure of the landscape, while the secondary
structure and dihedral consistency terms
$(\tilde{D}_{\mathrm{ss}},\tilde{D}_{\angle})$ inject finer-grained
gradients into local regions. Conceptually, this process reshapes the
quantum landscape: basins identified by quantum energies are refined by
superimposing the learned statistical potentials, thereby increasing
the resolution of the valleys where native-like conformations reside,
as illustrated schematically in Fig.~\ref{fig:energy_landscapes}c.

From the perspective of energy landscapes, the effect of fusion is to transform broad and shallow minima into narrower and more detailed wells. In the unfused quantum landscape, the lowest energy basin may contain a wide range of conformations with nearly indistinguishable energies, making it difficult to isolate the native like state. The inclusion of biological priors obtained from learning models introduces additional gradients within these basins, sharpening the surface so that conformations consistent with established secondary structure patterns are preferentially selected. The fused energy landscape therefore achieves higher effective resolution, guiding the search not only toward low energy regions but also toward structures that are biologically meaningful. This hybrid approach remains principled because it preserves the fundamental roles of both physics and data. Quantum algorithms ensure that the global form of the energy surface follows the laws of quantum mechanics, while learning based priors provide locally valid information extracted from empirical observations of protein structures. The fusion does not replace quantum mechanics but enriches it by adding resolution in regions where quantum methods are weakest. Likewise, the biological priors are not applied independently but are constrained and validated by the global quantum energy surface~\cite{pamidimukkala2024protein}. This interplay guarantees that the resulting predictions are both physically consistent and statistically credible.

In summary, this work demonstrates that near term quantum hardware, though limited in precision, can become practically useful when integrated with data driven models. The hybrid energy landscape serves as a conceptual bridge: quantum mechanics defines the valleys, while learning based refinement sharpens their contours. This principle extends beyond protein structure prediction to other problems where coarse quantum energy surfaces can be complemented by high resolution empirical potentials. The resulting framework is feasible with current quantum devices, grounded in physical theory, and capable of producing biologically interpretable structures.

\subsection*{Quantum Candidate Generation}
Our quantum prediction workflow builds upon our previous work~\cite{zhang2025prediction} and is fully executed on IBM's 127-qubit superconducting quantum processor hosted at the Cleveland Clinic. All reported structures are direct outputs generated by the quantum hardware after completion of the prediction pipeline, without reliance on classical post hoc substitution. The overall workflow proceeds as follows:

\paragraph{Amino-acid sequence encoding.}
The input to our framework is an amino-acid fragment of length $L$. 
We represent the sequence as $\mathrm{Str}=\{a_1,a_2,\dots,a_L\}$, 
and encode each residue into a set of Pauli operators acting on qubits~\cite{wang2024efficient}. 
A problem-specific Hamiltonian is then constructed to capture structural 
constraints, chirality, and residue interactions. In our implementation, 
the Hamiltonian takes the form
\[
H = H_{\mathrm{geom}} + H_{\mathrm{steric}} + H_{\mathrm{int}} + H_{\mathrm{misc}},
\]
where $H_{\mathrm{geom}}$ enforces backbone integrity, 
$H_{\mathrm{steric}}$ encodes steric exclusion and chirality, 
$H_{\mathrm{int}}$ models residue–residue interactions 
(We use Miyazawa--Jernigan potentials here), and $H_{\mathrm{misc}}$ 
covers additional geometric constraints. The Hamiltonian is expressed as a weighted sum of Pauli strings 
and mapped to qubits using string-based encodings.

\paragraph{Quantum circuit construction and VQE.}
We employ an ansatz-based Variational Quantum Eigensolver (VQE). 
Pauli terms are mapped into quantum circuits, and an EfficientSU2 ansatz 
with depth $d$ is used to parametrize the variational state. 
The quantum estimator evaluates
\[
E_q(\boldsymbol\theta) = \langle \psi(\boldsymbol\theta) \,|\, H \,|\, \psi(\boldsymbol\theta)\rangle,
\]
where $\boldsymbol\theta$ are tunable gate parameters. 
A classical optimizer iteratively updates $\boldsymbol\theta$ to minimize $E_q$, 
while the quantum sampler executes parameterized circuits on superconducting 
hardware (IBM System One, 127 qubits, Cleveland Clinic backend). 
The workflow includes compilation into hardware-native gates, 
job submission, and measurement of expectation values with finite precision.

\paragraph{Candidate extraction.}
After convergence (or a fixed number of iterations), measurement results 
are collected as bitstrings. These bitstrings are reverse-mapped into 
geometric vectors representing relative coordinates between consecutive residues. 
Post-processing yields $N$ low-energy candidate backbones 
$\{c_1,\dots,c_N\}$, each stored as a C$\alpha$-only coordinate set 
in \texttt{.xyz} format. These conformations represent low-energy predictions 
from the quantum model but do not directly include secondary-structure 
or dihedral features.

\subsection*{Deep-learning Feature Extraction}

\paragraph{Neural predictor.}
To complement the quantum outputs, we obtain structural features from 
NetSurfP-3.0, a state-of-the-art sequence-based predictor~\cite{hoie2022netsurfp}. 
Given the same amino-acid sequence, NetSurfP produces a residue-wise tabular 
output consisting of $19$ columns: SS8 probabilities ($8$), SS3 probabilities ($3$), 
disorder probabilities ($2$), relative solvent accessibility (RSA), absolute solvent 
accessibility (ASA), and predicted backbone dihedral angles $\phi,\psi$ in degrees~\cite{howe2025relationship}. 

\paragraph{Extracted features.}
For our framework, we utilize three categories of features:
\begin{itemize}
  \item $\mathbf{p}^{\,\mathrm{ss}}_i$: residue-wise SS3 or SS8 probabilities, normalized to sum to one.
  \item $(\phi_i,\psi_i)$: predicted backbone dihedral angles, converted from degrees to radians.
  \item $w_i$: RSA values in $[0,1]$, used as optional weights for angle-based consistency losses.
\end{itemize}
Missing or undefined entries (e.g., boundary residues) are imputed with uniform 
distributions or masked out. RSA values outside $[0,1]$ are clipped, and NaN values 
are replaced by $1.0$, ensuring robustness for downstream fusion.

\paragraph{Ramachandran-kernel induction of secondary-structure distributions.}
From $(\hat\phi_i,\hat\psi_i)$ we generate candidate secondary-structure 
distributions $\hat{\mathbf{p}}^{\,\mathrm{ss}}_i(c)$. 
We model the three principal Ramachandran basins using Gaussian kernels 
centered at $(\phi,\psi)=(-60^\circ,-45^\circ)$ (helix), 
$(-120^\circ,130^\circ)$ (strand), and $(0^\circ,0^\circ)$ (coil), 
with isotropic $\sigma=40^\circ$. 
The resulting scores are normalized to form SS3 probabilities. 
If SS8 is required, SS3 distributions are heuristically expanded into SS8 
via fixed mappings: H$\to$(H,G,I), E$\to$(E,B), and C$\to$(T,S,L)~\cite{ting2010neighbor}. 
This procedure augments each quantum candidate with interpretable 
secondary-structure features.

\paragraph{Outcome.}
At this stage, each candidate conformation $c$ is annotated with 
(i) its quantum energy $E_q(c)$, 
(ii) candidate dihedral angles $(\hat\phi_i(c),\hat\psi_i(c))$, and 
(iii) induced secondary-structure distributions 
$\hat{\mathbf{p}}^{\,\mathrm{ss}}_i(c)$. 
This feature-complete representation enables subsequent re-ranking 
through fusion with deep-learning features.

\subsection*{Energy-based Feature Fusion}

\paragraph{Notation and inputs.}
For a protein fragment of length $L$, the quantum stage produces up to $N$ candidate conformations 
$\mathcal{C}=\{c_1,\dots,c_N\}$, each associated with a quantum energy $E_q(c)\in\mathbb{R}$. 
In parallel, a neural predictor provides residue-wise priors, including 
secondary-structure probabilities $\mathbf{p}^{\,\mathrm{ss}}_i\in\Delta^{K-1}$ 
($K=3$ for SS3 or $K=8$ for SS8), backbone dihedral angles 
$\phi_i,\psi_i\in(-\pi,\pi]$, and optional residue weights $w_i\in[0,1]$ 
(derived from relative solvent accessibility; by default, $w_i\equiv 1$). 
Angles are represented in radians.

For each candidate $c$, we denote the corresponding dihedral estimates by 
$\hat\phi_i(c),\hat\psi_i(c)\in(-\pi,\pi]$. When explicit dihedrals are absent, 
we compute ``virtual'' dihedrals directly from C$\alpha$-only coordinates 
using the standard four-point torsion formula. From these angles, a residue-wise 
candidate SS distribution $\hat{\mathbf{p}}^{\,\mathrm{ss}}_i(c)$ is induced 
via Ramachandran kernels (see ``SS from angles'' below).

We define the fused energy of a candidate as
\begin{equation}
E_{\mathrm{fuse}}(c)\;=\;
\alpha\,\widetilde{E}_q(c)\;+\;
\beta\,\widetilde{D}_{\mathrm{ss}}(c)\;+\;
\gamma\,\widetilde{D}_{\angle}(c),
\label{eq:fusion}
\end{equation}
where $(\alpha,\beta,\gamma)>0$ are user-specified trade-off weights. 
Tildes indicate per-term normalization (\S\ref{subsec:norm}), ensuring 
all components are on comparable scales. Candidates are ranked by ascending 
$E_{\mathrm{fuse}}$.

\paragraph{Secondary-structure divergence.}
Let $\mathcal{I}_{\mathrm{ss}}\subseteq\{1,\dots,L\}$ denote residues for which both 
$\mathbf{p}^{\,\mathrm{ss}}_i$ and $\hat{\mathbf{p}}^{\,\mathrm{ss}}_i(c)$ are defined. 
For residue $i$, we measure the discrepancy between prior and candidate distributions 
using one of the following metrics:
\begin{equation}
d_{\mathrm{ss}}\!\left(\mathbf{p},\hat{\mathbf{p}}\right)=
\begin{cases}
-\displaystyle\sum_{k=1}^{K} p_k \log\big(\hat p_k+\varepsilon\big), & \text{Cross-Entropy (CE)},\\[6pt]
\displaystyle\sum_{k=1}^{K} p_k \log\frac{p_k+\varepsilon}{\hat p_k+\varepsilon}, & \text{Kullback--Leibler (KL)},\\[10pt]
\left\lVert \mathbf{p}-\hat{\mathbf{p}}\right\rVert_2, & \text{$L_2$ distance},
\end{cases}
\label{eq:ss_metric}
\end{equation}
with a small constant $\varepsilon>0$ for numerical stability. 
We then aggregate across residues by taking the mean:
\begin{equation}
D_{\mathrm{ss}}(c)\;=\;\frac{1}{|\mathcal{I}_{\mathrm{ss}}|}
\sum_{i\in\mathcal{I}_{\mathrm{ss}}} 
d_{\mathrm{ss}}\!\left(\mathbf{p}^{\,\mathrm{ss}}_i,\;\hat{\mathbf{p}}^{\,\mathrm{ss}}_i(c)\right).
\label{eq:ss_agg}
\end{equation}

\paragraph{Angle consistency.}
Let $\mathcal{I}_{\angle}\subseteq\{1,\dots,L\}$ be the set of residues with both 
predicted angles $(\phi_i,\psi_i)$ and candidate estimates $(\hat\phi_i(c),\hat\psi_i(c))$. 
To account for angular periodicity, we compute wrapped differences on the unit circle:
\begin{equation}
\Delta(\theta,\hat\theta)
=\mathrm{wrap}(\theta-\hat\theta)\in(-\pi,\pi],\qquad
d_{\mathbb{S}^1}^2(\theta,\hat\theta)=\Delta(\theta,\hat\theta)^2,
\label{eq:wrap}
\end{equation}
where $\mathrm{wrap}$ adds/subtracts multiples of $2\pi$ to map values into $(-\pi,\pi]$. 
For each residue, we accumulate $\phi$ and $\psi$ errors and compute an RSA-weighted mean:
\begin{equation}
D_{\angle}(c)\;=\;
\frac{
\displaystyle\sum_{i\in \mathcal{I}_{\angle}} 
w_i\bigl[\,d_{\mathbb{S}^1}^2\!\bigl(\phi_i,\hat\phi_i(c)\bigr)+
d_{\mathbb{S}^1}^2\!\bigl(\psi_i,\hat\psi_i(c)\bigr)\,\bigr]
}{
\displaystyle\sum_{i\in \mathcal{I}_{\angle}} w_i
},
\label{eq:angle_loss}
\end{equation}
where $w_i\in[0,1]$ if RSA weighting is applied, otherwise $w_i\equiv 1$. 
If $\sum_i w_i=0$, uniform weights are used.

\paragraph{Virtual dihedrals from C$\alpha$ coordinates.}
When only C$\alpha$ coordinates are available, we approximate dihedrals via the 
standard four-point formula. For consecutive points $p_0,p_1,p_2,p_3\in\mathbb{R}^3$, we define
\begin{align}
\mathbf{b}_0&=p_1-p_0,\quad 
\mathbf{b}_1=p_2-p_1,\quad 
\mathbf{b}_2=p_3-p_2,\nonumber\\
\hat{\mathbf{b}}_1&=\mathbf{b}_1/\|\mathbf{b}_1\|,\quad
\mathbf{v}=\mathbf{b}_0-\bigl(\mathbf{b}_0\!\cdot\!\hat{\mathbf{b}}_1\bigr)\hat{\mathbf{b}}_1,\quad
\mathbf{w}=\mathbf{b}_2-\bigl(\mathbf{b}_2\!\cdot\!\hat{\mathbf{b}}_1\bigr)\hat{\mathbf{b}}_1.\nonumber
\end{align}
The resulting dihedral is
\begin{equation}
\mathrm{dihedral}(p_0,p_1,p_2,p_3)\;=\;\mathrm{atan2}\!\left(
\bigl(\hat{\mathbf{b}}_1\times \mathbf{v}\bigr)\!\cdot\!\mathbf{w}\;,\;
\mathbf{v}\!\cdot\!\mathbf{w}
\right)\in(-\pi,\pi].
\label{eq:dihedral}
\end{equation}
This provides per-residue proxies for $\phi$ and $\psi$, while edge residues lacking 
four points are masked.

\paragraph{SS induction from angles.}
Given $(\hat\phi_i(c),\hat\psi_i(c))$, we infer SS3 probabilities 
$\hat{\mathbf{p}}^{\,\mathrm{ss3}}_i(c)\in\Delta^{2}$ using isotropic Gaussian kernels 
centered at canonical Ramachandran basins: helices (H) at $(-60^\circ,-45^\circ)$, 
strands (E) at $(-120^\circ,130^\circ)$, and coils (C) at $(0^\circ,0^\circ)$, 
with $\sigma=40^\circ$. 
After normalization,
\[
\hat{\mathbf{p}}^{\,\mathrm{ss3}}_i(c)
=\frac{1}{Z_i}\bigl[s_{i}^{(H)},\, s_{i}^{(E)},\, s_{i}^{(C)}\bigr],\quad
Z_i=\sum_{X\in\{H,E,C\}} s_{i}^{(X)}.
\]
If SS8 is required, we heuristically expand SS3 into SS8 by splitting H$\to$(H,G,I), 
E$\to$(E,B), and C$\to$(T,S,L) as
\begin{equation}
\hat{\mathbf{p}}^{\,\mathrm{ss8}}_i(c)=
\bigl[p_H\!\cdot\![0.8,0.1,0.1],\;
      p_E\!\cdot\![0.9,0.1],\;
      p_C\!\cdot\![\tfrac{1}{3},\tfrac{1}{3},\tfrac{1}{3}]\bigr],
\label{eq:ss8_expand}
\end{equation}
ordered as $[H,G,I,E,B,T,S,L]$ with $[p_H,p_E,p_C]=\hat{\mathbf{p}}^{\,\mathrm{ss3}}_i(c)$.

\paragraph{Per-term normalization.}\label{subsec:norm}
To ensure balanced contributions, each term is optionally min--max normalized across 
all candidates. For a vector $\mathbf{v}=(v_1,\dots,v_N)$,
\begin{equation}
\widetilde{v}_j=\begin{cases}
\dfrac{v_j - v_{\min}}{v_{\max}-v_{\min}}, & v_{\max}>v_{\min},\\[6pt]
0, & v_{\max}=v_{\min},
\end{cases}
\quad
v_{\min}=\min\nolimits_{\mathrm{finite}} v_j,\quad
v_{\max}=\max\nolimits_{\mathrm{finite}} v_j.
\label{eq:minmax}
\end{equation}
This is applied independently to $E_q$, $D_{\mathrm{ss}}$, and $D_{\angle}$, 
yielding $\widetilde{E}_q$, $\widetilde{D}_{\mathrm{ss}}$, and $\widetilde{D}_{\angle}$. 
Missing terms for a candidate are set to zero after normalization.

\paragraph{Final selection and re-ranking.}
For each candidate $c \in \mathcal{C}$, the fused energy $E_{\mathrm{fuse}}(c)$ is computed according to Eq.~\eqref{eq:fusion}. All candidates are then sorted in ascending order of $E_{\mathrm{fuse}}$, yielding a ranked list $\{c_{(1)}, c_{(2)}, \dots, c_{(N)}\}$ that reflects the balance between quantum energy and agreement with biological priors. To ensure numerical stability, all fused energies are normalized across candidates. In the case of identical fused values, preference is given to the structure with the lower raw quantum energy $E_q(c)$. The top-ranked structure $c_{(1)}$ is selected as the final prediction, while the next best candidates are retained for ensemble analysis or for comparison with independent evaluation metrics such as docking affinity. Each selected conformation is exported in both Cartesian (\texttt{.xyz}) and Protein Data Bank (\texttt{.pdb}) formats. A corresponding summary table records, for each candidate, its quantum energy, fused energy, and consistency scores for secondary structure and dihedral angles. This explicit ranking and export procedure ensures full traceability of the selection process and allows subsequent verification or reweighting under alternative parameter settings $(\alpha, \beta, \gamma)$.

\section*{Results}

\subsection*{Quantum Hardware Platform} 
All quantum experiments in this study were performed on IBM’s 127 qubit superconducting quantum processor located at the Cleveland Clinic~\cite{santos2016ibm}. This system is the first and, to date, the only quantum computer dedicated solely to biomedical research and pharmaceutical development. The processor employs a heavy hex lattice connectivity optimized for scalable two qubit operations. All computational tasks were executed through the IBM Quantum Runtime environment, which allows low latency interaction between classical and quantum computations. Standard error mitigation methods, including measurement error correction and zero noise extrapolation, were applied to minimize the influence of hardware noise. By utilizing this unique platform, our work demonstrates protein structure prediction on an actual quantum device designed specifically to advance research in biomedical and life science domains.

\subsection*{Overall Performance Comparison}

\begin{figure}[h]
  \centering
  \begin{subfigure}{0.48\linewidth}
    \centering
    \includegraphics[width=\linewidth]{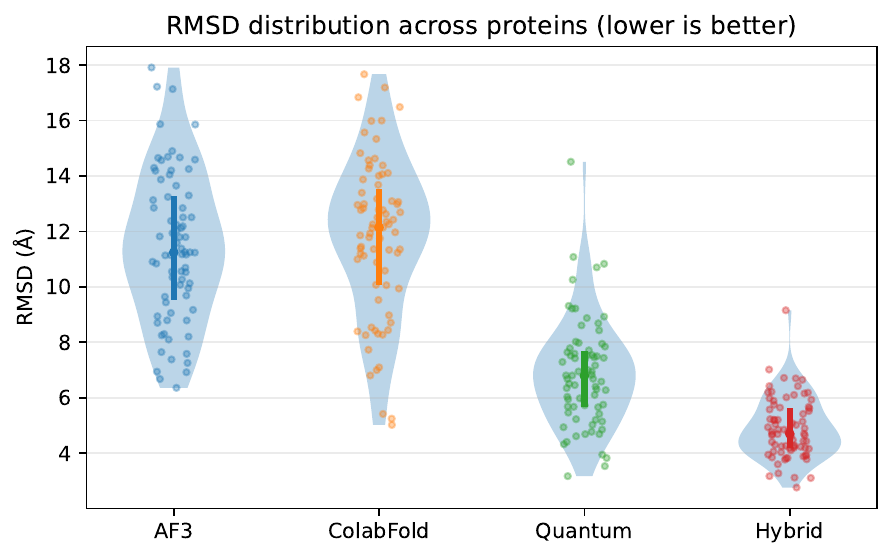}
    \caption{RMSD distribution across proteins for each method. Violin plots with jittered points show that Hybrid achieves the lowest RMSD overall, followed by Quantum, while AF3 and ColabFold exhibit higher values.}
    \label{fig:group:a}
  \end{subfigure}
  \hfill
  \begin{subfigure}{0.48\linewidth}
    \centering
    \includegraphics[width=\linewidth]{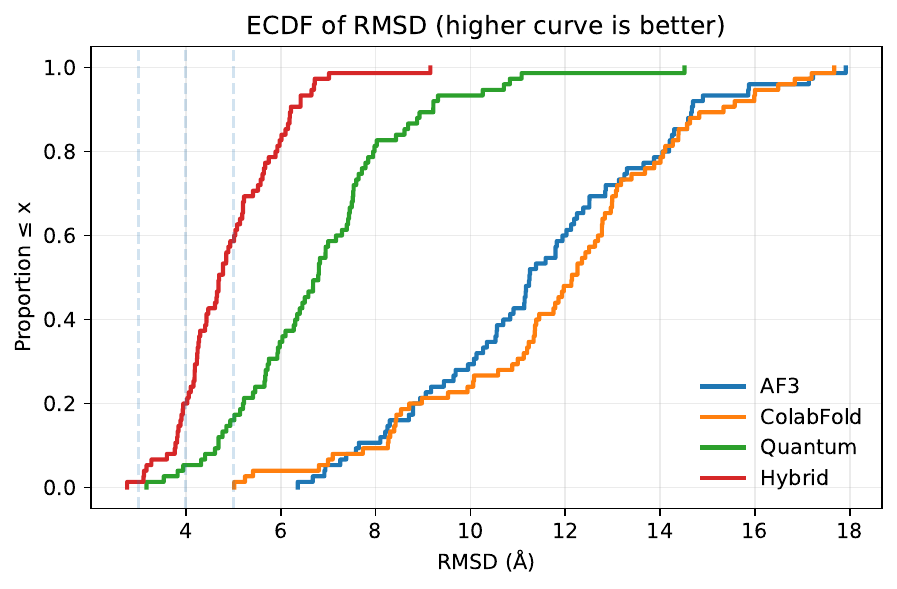}
    \caption{Empirical cumulative distribution function (ECDF) of RMSD. Higher curves indicate better performance. The Hybrid method dominates across nearly the entire RMSD range, followed by Quantum.}
    \label{fig:group:b}
  \end{subfigure}

  \vspace{0.5em}

  \begin{subfigure}{0.48\linewidth}
    \centering
    \includegraphics[width=\linewidth]{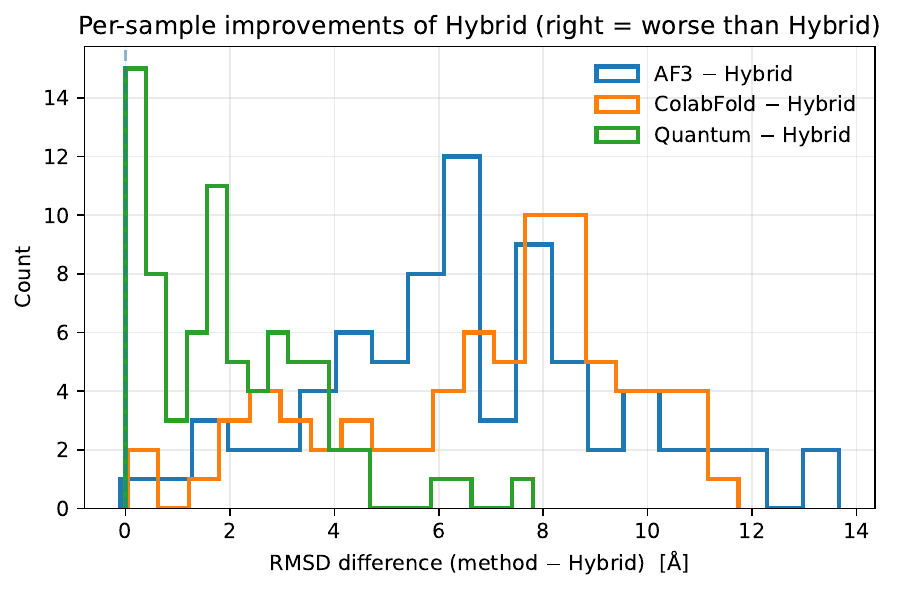}
    \caption{Per-sample improvements of Hybrid relative to other methods. Histogram of RMSD differences (method $-$ Hybrid) shows that Hybrid consistently outperforms AF3 and ColabFold, and provides further gains over Quantum for most proteins.}
    \label{fig:group:c}
  \end{subfigure}
  \hfill
  \begin{subfigure}{0.48\linewidth}
    \centering
    \includegraphics[width=\linewidth]{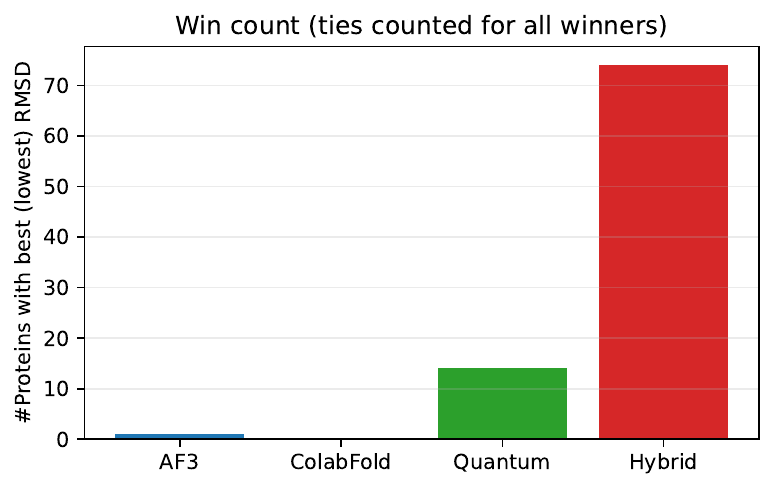}
    \caption{Win count comparison. Number of proteins where each method achieves the best (lowest) RMSD. Hybrid wins in the vast majority of cases, with Quantum in second place, while AF3 and ColabFold rarely achieve best performance.}
    \label{fig:group:d}
  \end{subfigure}

  \caption{Comprehensive comparison of AF3, ColabFold, Quantum, and Hybrid methods in terms of RMSD. Results demonstrate that the Hybrid framework consistently outperforms both classical AI-based approaches (AF3, ColabFold) and pure quantum predictions, achieving lower errors across distributions, cumulative statistics, per-sample improvements, and overall win counts.}
  \label{fig:group}
\end{figure}

We first evaluate backbone RMSD performance by comparing our hybrid method against three baselines: AlphaFold3 (AF3), ColabFold, and raw quantum-only predictions. Across 75 protein fragments from the PDBbind dataset~\cite{wang2005pdbbind}, our method consistently achieves the lowest RMSD values. Figure~\ref{fig:group} summarizes the RMSD distributions for all four approaches, and detailed statistics are provided in Table~\ref{tab:rmsd_summary}. Although AF3 and ColabFold attain reasonable accuracy on certain fragments, their median RMSDs remain substantially higher than those of the quantum-based methods. Specifically, AF3 yields a mean RMSD of $11.43$\,\AA\ (median $11.25$\,\AA) and ColabFold $11.79$\,\AA\ (median $12.14$\,\AA), both with large dispersions (standard deviations of $2.69$\,\AA\ and $2.84$\,\AA, respectively). By contrast, the quantum-only predictions markedly reduce the average error, with a mean RMSD of $6.85$\,\AA\ (median $6.79$\,\AA), moderate variability (standard deviation $1.92$\,\AA), and a worst-case deviation of $14.51$\,\AA. The hybrid re-ranking method further improves both accuracy and stability: it achieves the lowest mean RMSD of $4.89$\,\AA\ and median of $4.70$\,\AA, with a tighter distribution (standard deviation $1.10$\,\AA) and a maximum RMSD of $9.16$\,\AA, substantially smaller than those of AF3 ($17.92$\,\AA), ColabFold ($17.67$\,\AA), and the quantum-only baseline ($14.51$\,\AA). On average, this corresponds to RMSD reductions of $57.2\%$ relative to AF3, $58.5\%$ relative to ColabFold, and $28.6\%$ relative to quantum-only predictions, as summarized in Table~\ref{tab:rmsd_summary}.

\begin{table}[ht]
\centering
\caption{RMSD statistics across 75 protein fragments and average improvement of Hybrid relative to baseline methods.}
\label{tab:rmsd_summary}
\begin{tabular}{lcccccc}
\toprule
Method & Mean (\AA) & Median (\AA) & Std. Dev. (\AA) & Min (\AA) & Max (\AA) &  Improvement \\
\midrule
AF3 & 11.43 & 11.25 & 2.69 & 6.36 & 17.92 & 57.2\% \\
ColabFold & 11.79 & 12.14 & 2.84 & 5.02 & 17.67 & 58.5\% \\
Quantum-only & 6.85 & 6.79 & 1.92 & 3.17 & 14.51 & 28.6\% \\
Hybrid (ours) & 4.89 & 4.70 & 1.10 & 2.76 & 9.16 & -- \\
\bottomrule
\end{tabular}
\end{table}

To confirm the statistical robustness of these improvements, we performed paired one-tailed $t$-tests across all 75 fragments, testing the null hypothesis that the baseline RMSDs are not greater than those of the hybrid method. The results are summarized in Table~\ref{tab:ttest_summary}. All comparisons show extremely high statistical significance ($p \ll 0.001$) with large effect sizes (Cohen’s $d_z > 1$), indicating that the observed reductions in RMSD are both statistically and practically meaningful. The Wilcoxon signed-rank tests, included as a nonparametric robustness check, yield consistent one-tailed $p$-values below $10^{-10}$, further confirming the reliability of the improvements. These results demonstrate that the fused energy re-ranking scheme yields substantial and statistically significant accuracy gains over both deep learning and quantum-only baselines.

\begin{table}[ht]
\centering
\caption{Paired statistical comparison of RMSD differences (baseline $-$ hybrid) across 75 fragments. All $p$-values are one-tailed.}
\label{tab:ttest_summary}
\begin{tabular}{lccccc}
\toprule
Comparison & $n$ & Mean Diff. (\AA) & 95\% CI (\AA) & $t(74)$ & Cohen's $d_z$ \\
\midrule
AF3 vs Hybrid & 75 & 6.54 & [5.88, 7.19] & 19.93 & 2.30 \\
ColabFold vs Hybrid & 75 & 6.90 & [6.26, 7.54] & 21.50 & 2.48 \\
Quantum-only vs Hybrid & 75 & 1.95 & [1.57, 2.34] & 10.12 & 1.17 \\
\bottomrule
\end{tabular}
\\[4pt]
\small\textit{All comparisons show $p \ll 0.001$ (paired one-tailed $t$-test).}
\end{table}

As summarized in Figure~\ref{fig:group}, the Hybrid framework achieves superior accuracy across multiple evaluation perspectives. Specifically, the violin plots in panel~\ref{fig:group:a} show the distribution of RMSD values across proteins: Hybrid produces the tightest distribution with the lowest mean and variance, while Quantum-only predictions are moderately accurate, and AF3 and ColabFold display substantially higher errors. The empirical cumulative distribution function (ECDF) in panel~\ref{fig:group:b} further demonstrates this effect: the Hybrid curve lies consistently above the others, indicating that a larger proportion of fragments achieve low RMSD.  Panel~\ref{fig:group:c} examines per-sample improvements of Hybrid relative to the baselines by plotting histograms of RMSD differences (method $-$ Hybrid). The positive skew in these distributions highlights that Hybrid achieves lower RMSD in the vast majority of cases, not only compared to AF3 and ColabFold but also relative to Quantum-only predictions. Finally, the win-count analysis in panel~\ref{fig:group:d} quantifies how often each method achieves the best (lowest) RMSD for a given protein. Hybrid dominates this metric, winning in most cases, while Quantum-only predictions contribute a smaller number of wins, and AF3 and ColabFold rarely emerge as the best.  Together, these four complementary views confirm that the Hybrid approach provides consistent and statistically significant improvements over both classical and quantum-only baselines, delivering lower RMSD across distributions, cumulative statistics, per-sample comparisons, and best-case counts.

Per-fragment comparisons confirm these trends: in the majority of cases, Hybrid achieves the best RMSD, often reducing errors by several angstroms relative to AF3 and ColabFold. Even against the quantum-only baseline, Hybrid delivers consistent gains, with distributions shifted toward lower values and tighter clustering. Collectively, these results highlight that incorporating AI priors with quantum outputs via re-ranking is critical for achieving both accuracy and reliability in structure prediction. Table~\ref{tab:fused_rmsd_energy} reports per-fragment results for 75 protein fragments, including chain length, qubit count, and circuit depth. For each of the top-5 quantum candidates, we list RMSD, the fused score $E_{\mathrm{fuse}}$, and the raw quantum energy. Here, the \emph{Energy} column reflects the unadjusted eigenvalues obtained directly from the quantum simulation, whereas the \emph{Score} column incorporates post-processing terms (secondary-structure divergence and dihedral-angle consistency). This design allows direct comparison of the original quantum ranking versus the re-weighted fused ranking. 

The results confirm that fused scores provide better correlation with RMSD than raw 
quantum energies. For instance, in fragment \textbf{1dhj} (Len = 13, Depth = 373), the lowest fused score (0.05) corresponds to the most accurate candidate (RMSD = 3.94~\AA), while raw quantum energies alone would not distinguish among higher-RMSD candidates. Similarly, fragments such as \textbf{2fbq} and \textbf{3cd5} show near-degenerate raw energies across candidates, yet the fused scores effectively reorder them, aligning low scores with low RMSD. This demonstrates that the reweighting procedure is essential for guiding the selection toward physically accurate conformations beyond the quantum energies alone.

{\scriptsize
\setlength{\tabcolsep}{2pt}
\renewcommand{\arraystretch}{0.8}
\begin{longtable}{lrrrrrrrrrrrrrrrrrr}
\caption{Per-fragment summary with fused score (Score), RMSD, and quantum energy (integer) for the top-5 candidates. Left block reports chain length, qubit count, and circuit depth; each candidate block lists \textit{rmsd}, \textit{Score}, and \textit{Energy}.}
\label{tab:fused_rmsd_energy}\\
\toprule
\textbf{pdbid} & \textbf{Len} & \textbf{Qubits} & \textbf{Depth} &
\multicolumn{3}{c}{\textbf{Candidate1}} &
\multicolumn{3}{c}{\textbf{Candidate2}} &
\multicolumn{3}{c}{\textbf{Candidate3}} &
\multicolumn{3}{c}{\textbf{Candidate4}} &
\multicolumn{3}{c}{\textbf{Candidate5}} \\
\cmidrule(lr){5-7} \cmidrule(lr){8-10} \cmidrule(lr){11-13} \cmidrule(lr){14-16} \cmidrule(lr){17-19}
 &  &  &  &
\textit{rmsd} & \textit{Score} & \textit{Energy} &
\textit{rmsd} & \textit{Score} & \textit{Energy} &
\textit{rmsd} & \textit{Score} & \textit{Energy} &
\textit{rmsd} & \textit{Score} & \textit{Energy} &
\textit{rmsd} & \textit{Score} & \textit{Energy} \\
\midrule
\endfirsthead

\toprule
\textbf{pdbid} & \textbf{Len} & \textbf{Qubits} & \textbf{Depth} &
\multicolumn{3}{c}{\textbf{Candidate1}} &
\multicolumn{3}{c}{\textbf{Candidate2}} &
\multicolumn{3}{c}{\textbf{Candidate3}} &
\multicolumn{3}{c}{\textbf{Candidate4}} &
\multicolumn{3}{c}{\textbf{Candidate5}} \\
\cmidrule(lr){5-7} \cmidrule(lr){8-10} \cmidrule(lr){11-13} \cmidrule(lr){14-16} \cmidrule(lr){17-19}
 &  &  &  &
\textit{rmsd} & \textit{Score} & \textit{Energy} &
\textit{rmsd} & \textit{Score} & \textit{Energy} &
\textit{rmsd} & \textit{Score} & \textit{Energy} &
\textit{rmsd} & \textit{Score} & \textit{Energy} &
\textit{rmsd} & \textit{Score} & \textit{Energy} \\
\midrule
\endhead

\midrule
\multicolumn{19}{r}{\emph{Continued on next page}}\\
\midrule
\endfoot

\bottomrule
\endlastfoot

1bai & 11 & 72 & 293 & 8.43 & 0.55 & 6593 & 4.16 & 0.08 & 6632 & 5.60 & 1.43 & 6875 & 7.60 & 2.41 & 6920 & 7.93 & 2.95 & 7067 \\
1dhj & 13 & 92 & 373 & 3.94 & 0.05 & 15003 & 5.98 & 2.30 & 16124 & 5.76 & 1.34 & 16212 & 5.88 & 1.97 & 16507 & 8.43 & 2.49 & 16844 \\
1e1x & 14 & 102 & 413 & 5.08 & 0.69 & 23350 & 6.42 & 1.80 & 24170 & 6.70 & 2.30 & 24206 & 5.94 & 0.91 & 24359 & 9.03 & 2.79 & 24572 \\
1erb & 13 & 92 & 373 & 6.70 & 1.00 & 17154 & 8.88 & 1.35 & 17316 & 9.36 & 1.35 & 17364 & 9.90 & 1.44 & 17635 & 16.12 & 1.65 & 17751 \\
1jaq & 14 & 102 & 413 & 6.44 & 1.27 & 22849 & 4.40 & 0.13 & 22926 & 9.60 & 2.57 & 23195 & 6.98 & 1.74 & 23323 & 10.17 & 2.85 & 23451 \\
1k1y & 12 & 82 & 333 & 7.54 & 1.99 & 12498 & 4.06 & 0.57 & 13030 & 6.81 & 1.86 & 13240 & 8.10 & 2.64 & 13366 & 6.59 & 1.20 & 13434 \\
1yc4 & 13 & 92 & 373 & 7.59 & 1.53 & 16129 & 3.82 & 0.23 & 16216 & 3.89 & 0.73 & 16410 & 6.84 & 1.44 & 16544 & 8.50 & 3.00 & 16607 \\
1zsf & 10 & 63 & 257 & 3.10 & 0.52 & 4283 & 3.23 & 0.62 & 4288 & 3.82 & 1.37 & 4351 & 4.77 & 1.91 & 4371 & 4.96 & 2.19 & 4381 \\
2avo & 10 & 63 & 257 & 6.10 & 0.58 & 4711 & 9.65 & 1.33 & 4832 & 7.09 & 2.80 & 4838 & 6.29 & 0.92 & 4857 & 7.96 & 1.78 & 4869 \\
2bfq & 12 & 82 & 333 & 5.62 & 0.00 & 11785 & 6.59 & 1.77 & 12378 & 10.18 & 2.72 & 12746 & 6.15 & 1.31 & 12818 & 6.31 & 1.63 & 12953 \\
2bok & 10 & 63 & 257 & 5.95 & 1.01 & 4366 & 4.43 & 0.72 & 4383 & 6.35 & 2.26 & 4403 & 5.12 & 1.11 & 4503 & 5.80 & 1.24 & 4510 \\
2p4y & 13 & 92 & 373 & 5.48 & 1.75 & 17049 & 4.78 & 0.54 & 17138 & 5.75 & 2.03 & 17287 & 5.76 & 2.67 & 17364 & 5.44 & 1.28 & 17439 \\
2qbs & 11 & 72 & 293 & 5.14 & 0.20 & 6692 & 5.40 & 0.89 & 6895 & 6.87 & 2.17 & 6919 & 6.20 & 1.10 & 6928 & 6.80 & 3.00 & 7028 \\
2v7a & 12 & 82 & 333 & 7.74 & 1.86 & 12363 & 4.94 & 1.01 & 12554 & 6.06 & 1.27 & 12606 & 6.36 & 1.50 & 12638 & 6.87 & 2.83 & 12702 \\
2vwo & 10 & 63 & 257 & 3.84 & 1.00 & 4176 & 9.60 & 1.98 & 4360 & 7.53 & 1.85 & 4367 & 9.77 & 2.52 & 4380 & 5.49 & 1.13 & 4423 \\
2x4z & 12 & 82 & 333 & 5.89 & 1.54 & 11981 & 4.89 & 1.04 & 12245 & 6.95 & 1.75 & 12280 & 8.68 & 2.97 & 12387 & 6.62 & 1.68 & 12401 \\
2xxx & 12 & 82 & 333 & 5.41 & 0.82 & 14200 & 5.80 & 1.25 & 14623 & 5.57 & 1.13 & 14710 & 5.87 & 2.05 & 14732 & 6.78 & 3.00 & 14733 \\
2ygf & 13 & 92 & 373 & 7.65 & 1.84 & 15951 & 8.68 & 2.33 & 16277 & 4.69 & 0.75 & 16572 & 8.23 & 1.99 & 16692 & 5.83 & 1.78 & 16780 \\
3ag9 & 12 & 82 & 333 & 4.61 & 0.61 & 12638 & 5.97 & 1.16 & 12698 & 6.12 & 1.49 & 12724 & 6.60 & 2.04 & 12736 & 6.57 & 2.08 & 12840 \\
3b26 & 10 & 63 & 257 & 5.22 & 1.48 & 3769 & 3.12 & 0.95 & 3815 & 4.46 & 1.30 & 4014 & 3.53 & 0.98 & 4043 & 6.88 & 2.47 & 4066 \\
3bxh & 10 & 63 & 257 & 4.43 & 0.62 & 4109 & 6.80 & 1.96 & 4414 & 7.42 & 2.14 & 4418 & 6.70 & 1.61 & 4434 & 6.17 & 1.37 & 4458 \\
3cd5 & 11 & 72 & 293 & 4.63 & 1.00 & 6888 & 5.76 & 1.78 & 6896 & 4.28 & 0.87 & 6977 & 7.39 & 2.07 & 7026 & 6.78 & 1.90 & 7054 \\
3d7z & 14 & 102 & 413 & 11.08 & 1.81 & 22980 & 9.15 & 0.84 & 23231 & 14.27 & 2.32 & 23303 & 16.22 & 2.80 & 23515 & 9.51 & 1.07 & 23526 \\
3d83 & 10 & 63 & 257 & 7.24 & 1.77 & 4235 & 6.63 & 1.73 & 4273 & 7.18 & 2.81 & 4368 & 10.83 & 2.74 & 4389 & 4.85 & 1.00 & 4390 \\
3f16 & 14 & 102 & 413 & 6.51 & 1.24 & 22987 & 7.41 & 1.65 & 23096 & 7.13 & 1.35 & 23126 & 9.65 & 2.55 & 23221 & 6.01 & 1.00 & 23412 \\
3fwv & 10 & 63 & 257 & 5.30 & 2.00 & 4098 & 2.76 & 0.97 & 4264 & 5.00 & 1.55 & 4351 & 4.55 & 1.14 & 4360 & 4.32 & 1.07 & 4393 \\
3gnw & 14 & 102 & 413 & 3.93 & 0.21 & 21203 & 7.41 & 1.31 & 21721 & 8.61 & 2.75 & 21920 & 8.32 & 1.90 & 21950 & 6.22 & 1.08 & 22162 \\
3jrx & 14 & 102 & 413 & 5.46 & 0.70 & 22813 & 3.82 & 0.38 & 23166 & 6.39 & 1.30 & 23362 & 7.52 & 1.30 & 23374 & 7.89 & 3.00 & 23733 \\
3n2v & 14 & 102 & 413 & 4.85 & 0.48 & 21831 & 6.94 & 1.64 & 22282 & 5.56 & 0.87 & 22306 & 8.78 & 2.22 & 22449 & 8.48 & 2.82 & 22620 \\
3tfp & 14 & 102 & 413 & 5.20 & 0.12 & 20779 & 5.72 & 0.59 & 21912 & 10.12 & 2.59 & 22110 & 6.05 & 1.45 & 22956 & 6.25 & 1.63 & 23026 \\
3vf7 & 10 & 63 & 257 & 4.23 & 1.34 & 3975 & 6.21 & 1.36 & 4298 & 7.95 & 1.86 & 4330 & 7.64 & 1.44 & 4356 & 7.89 & 1.71 & 4386 \\
3vje & 14 & 102 & 413 & 4.70 & 0.36 & 22112 & 5.14 & 1.60 & 23477 & 4.73 & 0.87 & 23629 & 7.09 & 2.99 & 23848 & 5.77 & 1.66 & 23858 \\
3w5n & 11 & 72 & 293 & 4.68 & 1.68 & 7824 & 5.53 & 2.04 & 7943 & 4.65 & 0.67 & 8012 & 6.39 & 2.21 & 8084 & 7.06 & 2.67 & 8103 \\
4acc & 13 & 92 & 373 & 5.23 & 0.14 & 15748 & 6.70 & 2.00 & 15795 & 6.99 & 2.65 & 16160 & 6.93 & 2.01 & 16245 & 6.68 & 1.30 & 16379 \\
4aoi & 14 & 102 & 413 & 7.81 & 1.36 & 23245 & 4.24 & 0.56 & 23375 & 6.52 & 0.62 & 23832 & 12.05 & 2.60 & 23916 & 10.70 & 2.33 & 24366 \\
4b6p & 12 & 82 & 333 & 5.20 & 0.18 & 12711 & 6.94 & 1.62 & 12919 & 8.31 & 2.20 & 12973 & 6.92 & 1.55 & 13098 & 8.89 & 2.27 & 13181 \\
4ceb & 14 & 102 & 413 & 8.12 & 1.98 & 22813 & 7.63 & 1.21 & 23574 & 7.98 & 1.95 & 23719 & 6.15 & 0.67 & 23800 & 8.45 & 2.97 & 24288 \\
4cig & 14 & 102 & 413 & 4.03 & 0.50 & 21376 & 8.48 & 2.06 & 22277 & 8.02 & 2.87 & 22640 & 5.78 & 0.95 & 22756 & 6.80 & 1.88 & 22835 \\
4cjr & 14 & 102 & 413 & 7.43 & 2.00 & 23168 & 6.40 & 1.65 & 23770 & 6.69 & 1.75 & 23783 & 5.89 & 0.90 & 23873 & 7.63 & 2.77 & 23949 \\
4clj & 14 & 102 & 413 & 5.67 & 0.24 & 23969 & 10.44 & 1.76 & 24011 & 9.04 & 1.20 & 24024 & 9.93 & 1.69 & 24126 & 9.60 & 1.28 & 24373 \\
4dv8 & 14 & 102 & 413 & 8.93 & 1.66 & 23407 & 10.20 & 1.98 & 23494 & 5.19 & 0.72 & 23609 & 8.44 & 1.46 & 23816 & 5.24 & 1.32 & 23910 \\
4dzy & 13 & 92 & 373 & 5.64 & 0.31 & 17531 & 8.27 & 1.31 & 17961 & 10.26 & 1.96 & 17975 & 5.75 & 0.96 & 18008 & 11.48 & 2.63 & 18026 \\
4eor & 13 & 92 & 373 & 5.75 & 0.00 & 16251 & 6.25 & 0.81 & 16301 & 7.78 & 2.81 & 16691 & 7.62 & 1.91 & 16739 & 7.64 & 1.99 & 16794 \\
4f5y & 11 & 72 & 293 & 5.27 & 1.11 & 6408 & 4.69 & 0.73 & 6701 & 6.94 & 2.28 & 6712 & 7.30 & 2.30 & 6784 & 7.45 & 2.73 & 6811 \\
4fp1 & 14 & 102 & 413 & 6.19 & 0.86 & 22564 & 7.51 & 1.06 & 23026 & 8.46 & 1.64 & 23516 & 9.18 & 1.72 & 23561 & 9.31 & 3.00 & 23621 \\
4gih & 12 & 82 & 333 & 8.61 & 1.78 & 12586 & 5.58 & 0.21 & 12586 & 7.79 & 1.09 & 12699 & 7.94 & 2.17 & 12757 & 9.60 & 1.21 & 12817 \\
4jpx & 13 & 92 & 373 & 4.30 & 0.03 & 16962 & 5.22 & 0.10 & 16985 & 5.47 & 0.62 & 17154 & 7.71 & 2.39 & 17207 & 7.56 & 1.63 & 17594 \\
4jpy & 14 & 102 & 413 & 6.58 & 0.60 & 23332 & 5.98 & 0.35 & 23448 & 7.28 & 2.60 & 24200 & 6.99 & 2.34 & 24266 & 10.17 & 3.00 & 24271 \\
4mc1 & 10 & 63 & 257 & 4.19 & 0.00 & 4092 & 6.28 & 1.53 & 4231 & 7.52 & 2.50 & 4262 & 7.17 & 2.24 & 4338 & 8.67 & 2.70 & 4347 \\
4oc5 & 12 & 82 & 333 & 6.42 & 0.13 & 13634 & 6.93 & 1.17 & 13734 & 8.02 & 1.26 & 13780 & 7.99 & 2.75 & 13819 & 7.58 & 1.23 & 13881 \\
4qrh & 14 & 102 & 413 & 5.30 & 1.48 & 22858 & 4.90 & 1.34 & 22869 & 7.50 & 2.00 & 23153 & 3.76 & 0.89 & 23606 & 6.36 & 1.51 & 23701 \\
4tmk & 14 & 102 & 413 & 5.52 & 1.03 & 22590 & 4.25 & 0.21 & 22620 & 9.16 & 2.42 & 22655 & 5.00 & 0.95 & 22696 & 6.27 & 1.17 & 22744 \\
4urz & 12 & 82 & 333 & 6.97 & 2.00 & 11360 & 5.92 & 1.78 & 11657 & 4.18 & 0.92 & 11984 & 8.07 & 2.05 & 12075 & 6.79 & 1.34 & 12126 \\
4zbf & 14 & 102 & 413 & 5.46 & 2.00 & 23122 & 8.65 & 2.24 & 23773 & 8.73 & 1.60 & 23828 & 3.78 & 1.11 & 23893 & 5.27 & 1.40 & 23930 \\
4zgk & 13 & 92 & 373 & 6.00 & 0.50 & 16218 & 3.90 & 0.07 & 16278 & 7.29 & 1.85 & 16758 & 7.20 & 1.05 & 16916 & 7.66 & 3.00 & 17045 \\
5c2o & 14 & 102 & 413 & 7.02 & 0.94 & 22601 & 7.41 & 1.00 & 22866 & 9.22 & 1.14 & 22933 & 9.45 & 1.16 & 23186 & 9.80 & 3.00 & 23525 \\
5cqu & 13 & 92 & 373 & 5.04 & 0.00 & 17865 & 5.68 & 1.07 & 17957 & 7.72 & 2.39 & 18041 & 6.24 & 2.02 & 18241 & 6.91 & 2.02 & 18313 \\
5cxa & 11 & 72 & 293 & 3.60 & 0.00 & 6946 & 5.08 & 1.66 & 7060 & 5.02 & 1.46 & 7126 & 7.98 & 1.79 & 7154 & 8.47 & 3.00 & 7177 \\
5g2g & 10 & 63 & 257 & 3.27 & 0.05 & 3931 & 9.56 & 2.44 & 3978 & 4.46 & 1.13 & 3981 & 6.90 & 2.24 & 4008 & 6.03 & 1.83 & 4018 \\
5kqx & 10 & 63 & 257 & 5.72 & 1.67 & 4337 & 4.47 & 0.61 & 4513 & 5.93 & 1.79 & 4540 & 6.09 & 1.81 & 4587 & 6.82 & 3.00 & 4627 \\
5kr2 & 10 & 63 & 257 & 8.14 & 1.39 & 4114 & 4.26 & 0.33 & 4129 & 6.01 & 0.60 & 4219 & 6.68 & 1.21 & 4326 & 8.36 & 3.00 & 4337 \\
5nkb & 14 & 102 & 413 & 6.42 & 0.21 & 22571 & 7.68 & 2.10 & 23209 & 7.00 & 1.57 & 23312 & 8.53 & 2.14 & 23379 & 12.41 & 2.47 & 23391 \\
5nkc & 12 & 82 & 333 & 6.72 & 0.00 & 12920 & 8.40 & 1.34 & 13383 & 12.79 & 1.73 & 13451 & 14.51 & 2.37 & 13609 & 9.40 & 3.00 & 13688 \\
5nkd & 11 & 72 & 293 & 7.57 & 2.00 & 7193 & 6.98 & 1.56 & 7436 & 4.10 & 0.89 & 7482 & 5.66 & 0.98 & 7513 & 6.72 & 1.00 & 7917 \\
5otz & 13 & 92 & 373 & 4.78 & 0.20 & 16088 & 7.46 & 0.72 & 16246 & 7.62 & 0.91 & 16292 & 8.63 & 1.57 & 16337 & 9.37 & 3.00 & 16564 \\
5tp0 & 10 & 63 & 257 & 4.64 & 1.00 & 4199 & 5.46 & 1.12 & 4243 & 6.38 & 1.17 & 4459 & 7.16 & 1.81 & 4543 & 8.73 & 2.40 & 4559 \\
6ezq & 10 & 63 & 257 & 3.17 & 0.00 & 4179 & 4.34 & 0.41 & 4226 & 5.82 & 2.92 & 4302 & 4.54 & 2.50 & 4307 & 4.37 & 1.70 & 4313 \\
6g3a & 12 & 82 & 333 & 6.95 & 1.31 & 11268 & 6.68 & 1.24 & 12247 & 7.76 & 2.50 & 12521 & 5.52 & 1.09 & 12728 & 7.24 & 1.64 & 12737 \\
6g98 & 11 & 72 & 293 & 7.16 & 1.87 & 7254 & 6.22 & 1.12 & 7394 & 7.44 & 2.36 & 7531 & 7.08 & 1.77 & 7604 & 7.29 & 2.00 & 7639 \\
6i63 & 13 & 92 & 373 & 4.18 & 0.00 & 18033 & 5.42 & 0.62 & 18229 & 5.81 & 1.74 & 18359 & 6.14 & 2.95 & 18659 & 5.98 & 2.39 & 18694 \\
6o94 & 14 & 102 & 413 & 13.25 & 1.16 & 23708 & 6.18 & 0.75 & 24734 & 8.41 & 0.96 & 25334 & 10.22 & 2.98 & 25375 & 9.22 & 2.63 & 25403 \\
6o95 & 14 & 102 & 413 & 5.93 & 0.62 & 22550 & 6.38 & 0.76 & 22591 & 8.99 & 1.82 & 23156 & 10.22 & 2.97 & 23315 & 8.06 & 1.18 & 23337 \\
6prf & 10 & 63 & 257 & 3.88 & 0.81 & 4056 & 8.09 & 0.90 & 4108 & 4.05 & 1.93 & 4168 & 6.83 & 0.97 & 4203 & 8.89 & 3.00 & 4207 \\
6qas & 11 & 72 & 293 & 6.65 & 1.15 & 6729 & 7.84 & 1.49 & 7231 & 10.67 & 1.95 & 7279 & 12.14 & 2.14 & 7297 & 8.82 & 1.76 & 7308 \\
6udv & 14 & 102 & 413 & 5.02 & 0.00 & 24186 & 5.62 & 1.10 & 24738 & 6.83 & 2.61 & 24867 & 6.21 & 2.02 & 24882 & 6.46 & 2.30 & 25107 \\
\end{longtable}
}

\subsection*{Case Study: Energy Decomposition}

\begin{figure}[t]
  \centering
  \begin{subfigure}{0.48\linewidth}
    \centering
    \includegraphics[width=\linewidth]{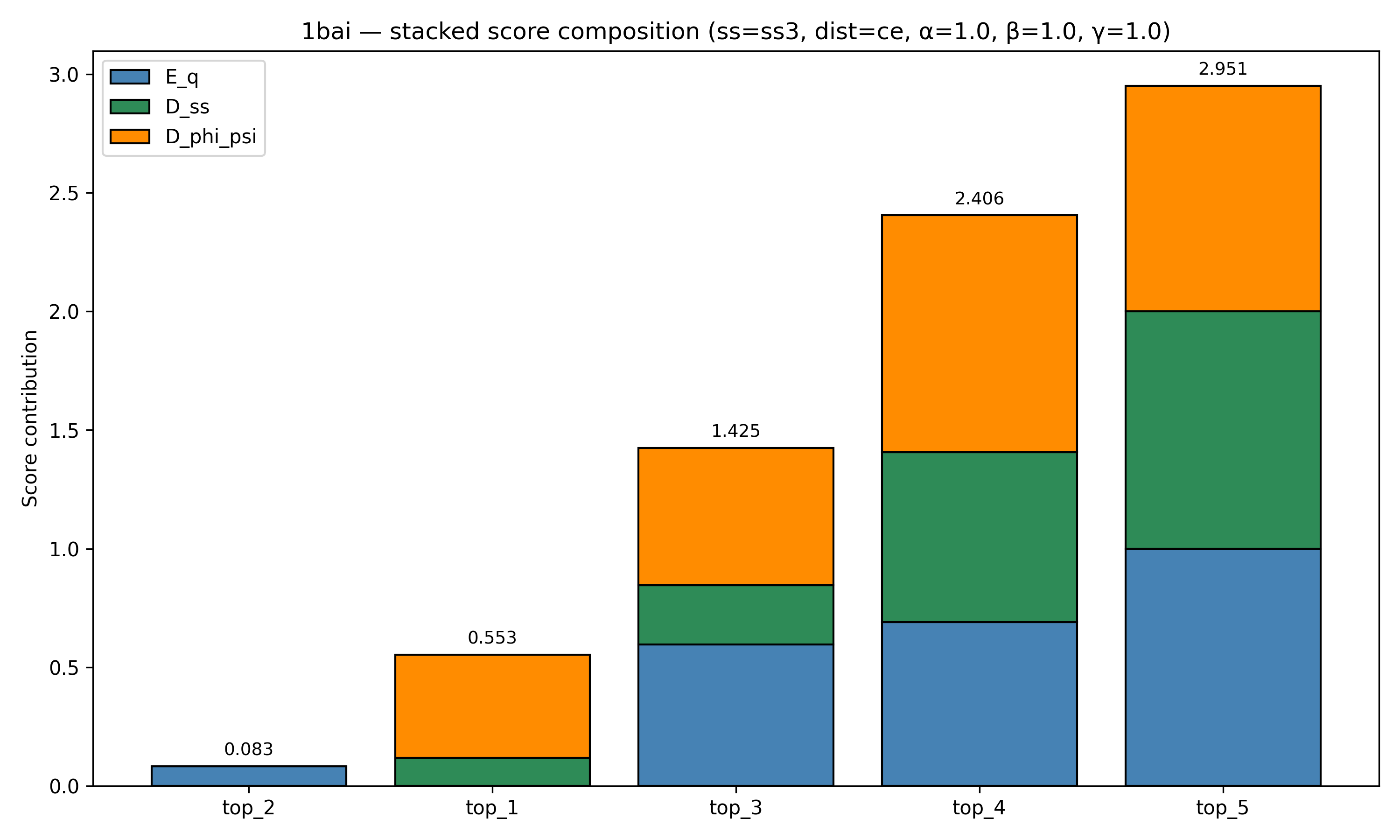}
    \caption{Case 1: PDB ID 1bai. Quantum, secondary structure, and torsional angle terms all contribute significantly to ranking.}
    \label{fig:case:1bai}
  \end{subfigure}
  \hfill
  \begin{subfigure}{0.48\linewidth}
    \centering
    \includegraphics[width=\linewidth]{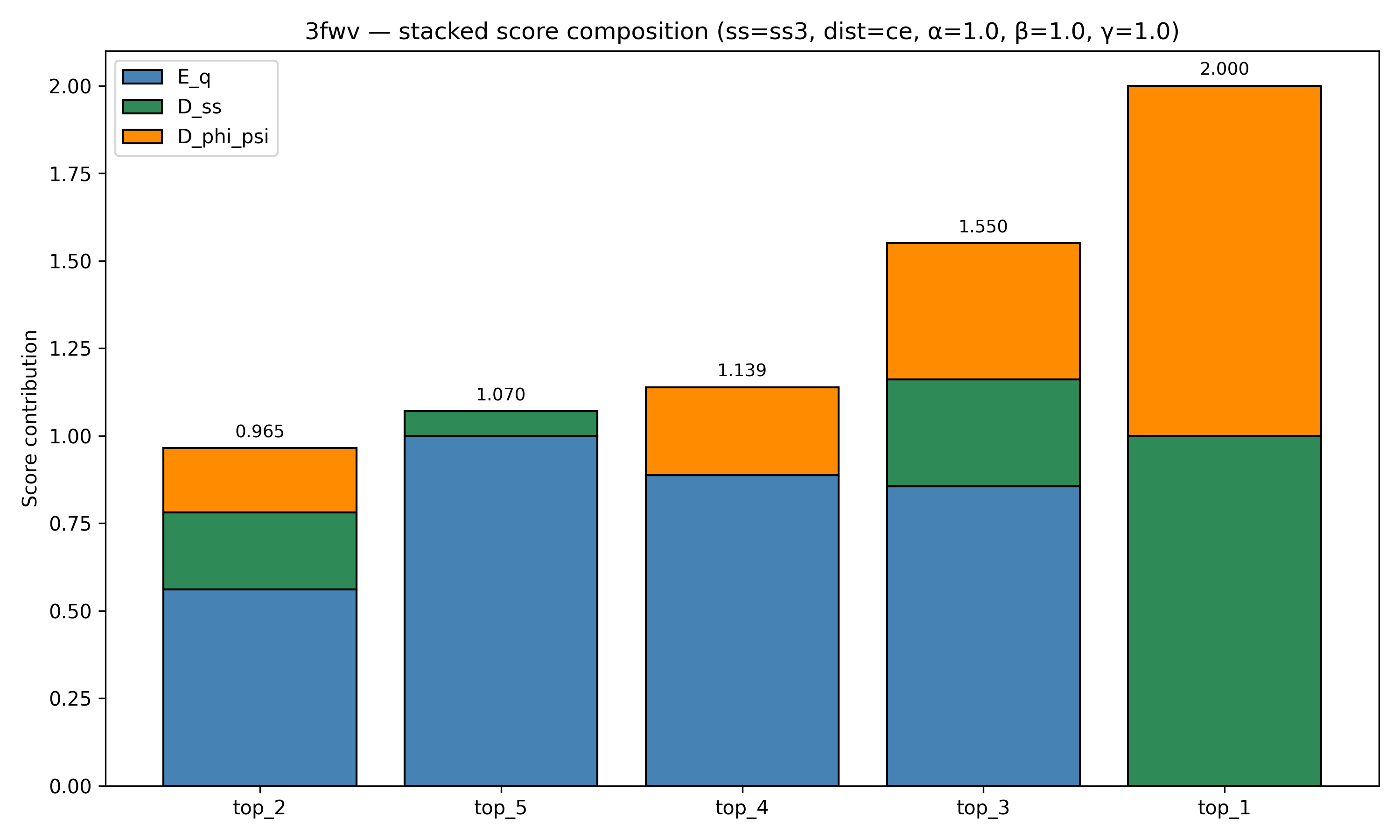}
    \caption{Case 2: PDB ID 3fwv. Here, quantum energy dominates, but secondary structure and torsional terms refine the final choice.}
    \label{fig:case:3fwv}
  \end{subfigure}

  \vspace{0.5em}

  \begin{subfigure}{0.48\linewidth}
    \centering
    \includegraphics[width=\linewidth]{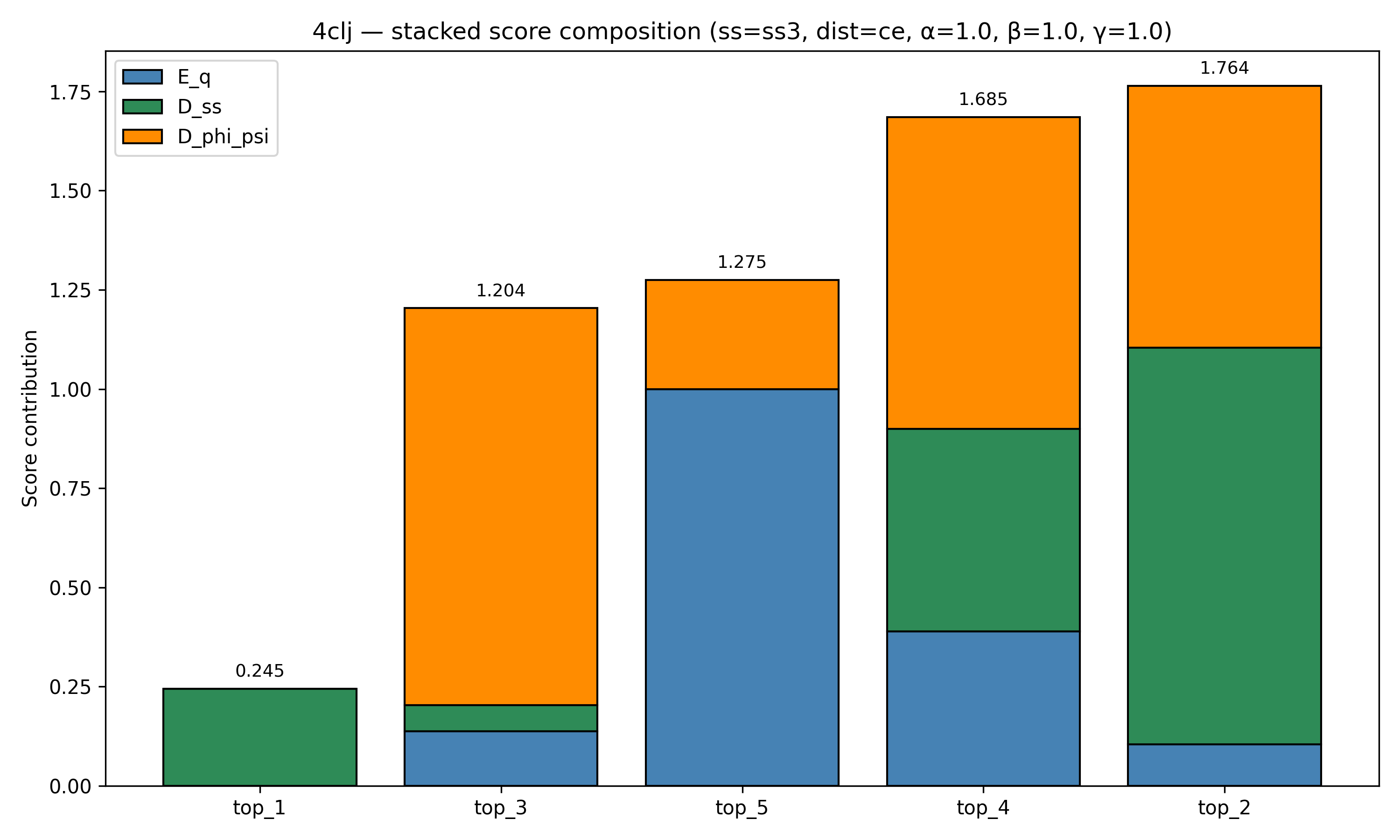}
    \caption{Case 3: PDB ID 4clj. Torsional angle differences provide the largest discriminative power among candidates.}
    \label{fig:case:4clj}
  \end{subfigure}
  \hfill
  \begin{subfigure}{0.48\linewidth}
    \centering
    \includegraphics[width=\linewidth]{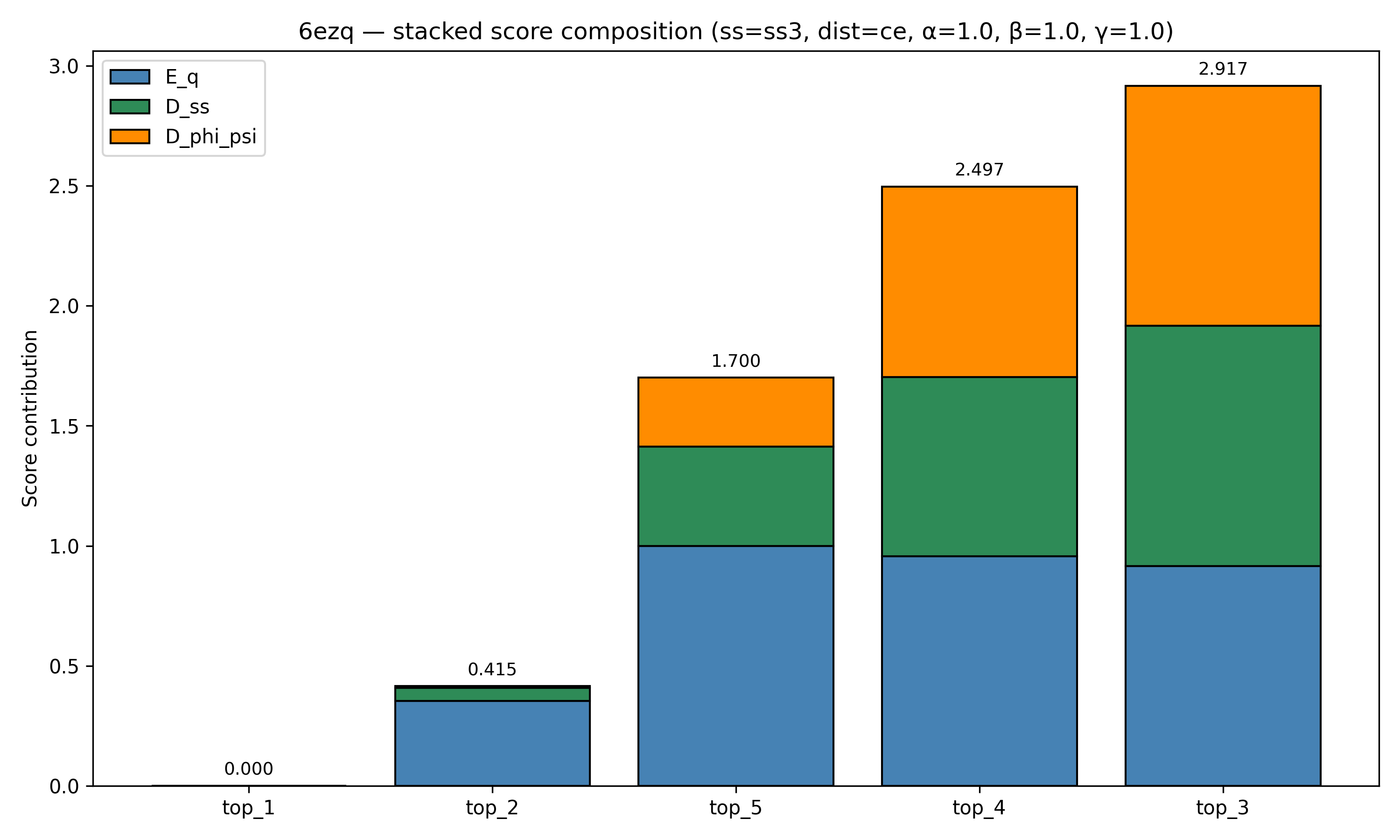}
    \caption{Case 4: PDB ID 6ezq. A balanced contribution from all three terms leads to robust re-ranking.}
    \label{fig:case:6ezq}
  \end{subfigure}

  \caption{Case study of score composition in four representative protein fragments. Each stacked bar plot shows contributions from quantum energy ($E_q$, blue), secondary-structure distribution difference ($D_{ss}$, green), and torsional angle difference ($D_{\phi\psi}$, orange). These examples highlight that all three terms are necessary and complementary: while quantum energy provides a baseline signal, structural priors refine and stabilize the ranking, ensuring that the hybrid method consistently selects the most accurate conformation.}
  \label{fig:case_study}
\end{figure}

To further illustrate the mechanism and rationality of our hybrid re-ranking framework, we present a case study on four representative protein fragments (PDB IDs: 1bai, 3fwv, 4clj, and 6ezq). For each case, Figure~\ref{fig:case_study} shows the stacked contributions of the three energy terms included in our scoring function: quantum energy ($E_q$), secondary-structure distribution difference ($D_{ss}$), 
and torsional angle difference ($D_{\phi\psi}$). These examples highlight how the hybrid scheme integrates heterogeneous sources of information to robustly discriminate between candidate conformations.

In the case of \textbf{1bai}, all three terms contribute substantially to the re-ranking score. Although the raw quantum energy already provides a strong signal, the inclusion of $D_{ss}$ and $D_{\phi\psi}$ further separates near-optimal candidates from suboptimal ones. This demonstrates that secondary-structure and torsional priors refine the decision boundary, preventing over-reliance on quantum estimates alone.

For \textbf{3fwv}, the stacked decomposition reveals that $E_q$ dominates the contribution. However, relying solely on $E_q$ would lead to ambiguities among close-energy candidates. Here, both $D_{ss}$ and $D_{\phi\psi}$ provide additional discriminative power that corrects for minor mis-rankings, ensuring that the most physically plausible structure is selected.

The case of \textbf{4clj} is particularly illustrative: the torsional angle term $D_{\phi\psi}$ clearly provides the largest contribution to the re-ranking score. This reflects the fact that torsional angles encode fine-grained backbone geometry, which cannot be fully captured by quantum or secondary-structure terms alone. In this scenario, incorporating angular information proves critical for identifying the correct conformation.

Finally, \textbf{6ezq} demonstrates a balanced contribution from all three components. While $E_q$ anchors the candidate energies, $D_{ss}$ and $D_{\phi\psi}$ jointly refine the selection, leading to a robust consensus ranking. The outcome highlights the complementary nature of the terms: none of them alone is sufficient to guarantee robust selection, but together they provide a consistent and reliable signal.

Across these diverse cases, we observe that the hybrid method avoids overfitting to a single source of information and leverages the orthogonal strengths of quantum energies, secondary structure consistency, and torsional priors. This decomposition analysis thus provides strong evidence that each term in our scoring function is necessary and that their integration yields a principled and interpretable framework for structure selection.

\subsection*{Correlation between Fused Energy and Structural Accuracy}

\begin{figure}[t]
    \centering
    \includegraphics[width=0.8\textwidth]{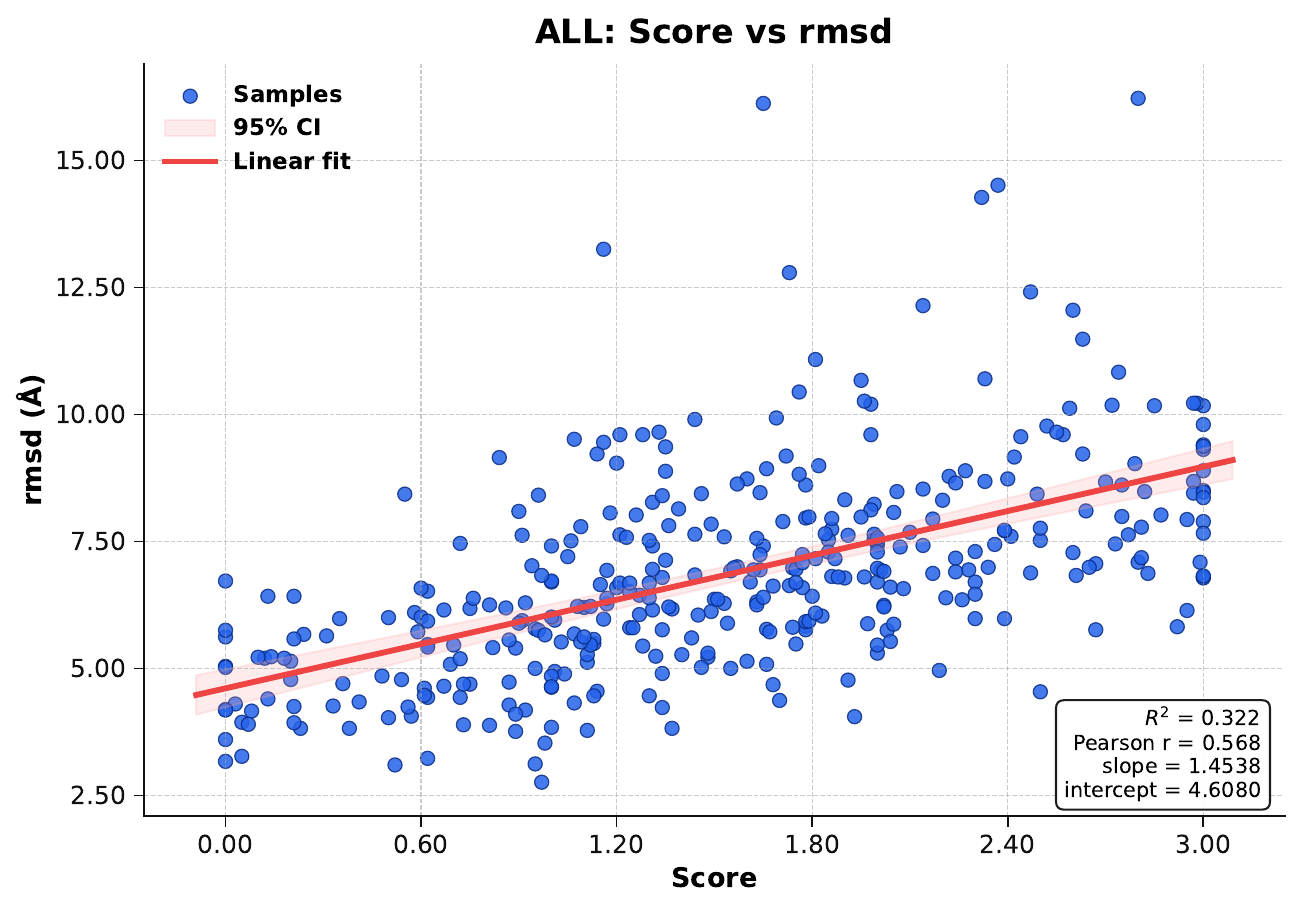}
    \caption{
    Correlation between fused energy scores and RMSD across all 375 candidate 
    structures (75 protein fragments, 5 conformations each). 
    Lower $E_{\mathrm{fuse}}$ values are associated with smaller RMSD, 
    confirming that the proposed energy reweighting scheme effectively 
    discriminates near-native conformations. 
    }
    \label{fig:correlation}
\end{figure}

To evaluate whether the fused energy function $E_{\mathrm{fuse}}$ effectively discriminates high-quality from low-quality conformations, we analyzed the correlation between $E_{\mathrm{fuse}}$ scores and RMSD values across all 375 candidate structures (75 fragments $\times$ top-5 conformations). Figure~\ref{fig:correlation} shows the scatter plot of RMSD versus fused energy scores, with a fitted regression line and 95\% confidence interval. 

The results reveal a clear positive correlation ($R^2 = 0.322$, Pearson $r = 0.568$), indicating that conformations with lower fused energy are generally associated with smaller RMSD values. This demonstrates that the fused energy function is capable of ranking candidate structures in a manner consistent with their structural accuracy. Importantly, the correlation is observed not only for the best-selected conformations but also across all five candidates per fragment, confirming that the energy reweighting scheme systematically guides the re-ranking process toward biologically meaningful conformations. These findings validate the design of our scoring function and its effectiveness in bridging quantum physical predictions with deep-learning priors.

\subsection*{Scalability and Robustness}

A central requirement for advancing quantum algorithms in protein structure prediction is demonstrating not only accuracy on near-term systems, but also pathways to scalability and robustness. Our framework addresses these considerations along two complementary dimensions.

\textbf{Scalability.}  
The encoding strategy employed here naturally generalizes to larger protein fragments and related molecular systems. The Hamiltonian construction based on backbone torsions scales linearly in qubit count with residue number, while coarse-grained contact models allow quadratic scaling with controllable sparsity. Although current experiments were restricted to fragments of up to 14 residues due to qubit limitations, the modularity of the encoding enables systematic extension: longer sequences can be partitioned into overlapping fragments, solved independently, and stitched through geometric consistency constraints. Moreover, the concept of energy fusion is not restricted to proteins; any molecular system that combines a physics-based Hamiltonian with learned statistical potentials (e.g., RNA folding, ligand docking, or peptide–membrane interactions) can be integrated into the same workflow. This demonstrates that the proposed method offers a blueprint for scalable hybrid quantum–classical modeling beyond the protein-folding domain.

\textbf{Robustness.}  
Noisy intermediate-scale quantum (NISQ) hardware imposes challenges of decoherence, limited connectivity, and sampling noise. Our design incorporates multiple mechanisms to ensure robustness against these effects. Variational ansätze are hardware-efficient which is EfficientSU2, respecting the device coupling graph and minimizing SWAP overhead~\cite{choudhary2021quantum}. Measurements are grouped into commuting sets with adaptive shot allocation to reduce variance, and error mitigation strategies such as zero-noise extrapolation and symmetry verification are applied to counter systematic biases. Crucially, the deep-learning priors serve not only to refine structural features but also to stabilize quantum outputs against hardware fluctuations: even when raw quantum energies exhibit noise-induced variance, the fused energy landscape remains discriminative, improving reliability across independent runs.

\textbf{Implications.}  
Together, scalability and robustness highlight the quantum contribution of this work. By introducing energy fusion as a modular principle, the framework provides a general recipe for incorporating domain-informed potentials into quantum optimization, enabling broader applicability across molecular modeling tasks. By demonstrating noise-tolerant execution on real 127-qubit hardware, it establishes a concrete benchmark for how hybrid quantum–classical strategies can be engineered today while providing a pathway toward utility-level quantum computing in the future.

\section*{Discussion}

The results of this study show that hybrid quantum and classical strategies can achieve practical utility in protein structure prediction, even with the limitations of current noisy intermediate scale quantum hardware. By integrating physically grounded variational quantum eigensolver (VQE) predictions with learning based priors derived from large structural datasets, our framework consistently generates conformations that are both energetically favorable and biologically interpretable. This outcome emphasizes the value of combining first principles quantum mechanics with data derived learning to overcome the limitations of each approach when used alone. Compared with leading deep learning models such as AlphaFold3 and ColabFold, our hybrid framework achieves lower RMSD values across a benchmark of seventy five fragments. The improvement is most evident in short or flexible sequences where purely data driven models often extrapolate poorly. At the same time, pure quantum predictions, though physically meaningful, lack essential features such as secondary structure motifs and consistent dihedral angles. The hybrid workflow addresses these gaps by using learning based priors to guide the ranking of quantum candidates. This demonstrates a complementary relationship in which quantum computation ensures physical fidelity, while deep learning contributes structural completeness and biological realism.

Our findings highlight a central principle for applying quantum algorithms to complex biological systems: quantum hardware alone is not yet capable of producing feature rich predictions, but when combined with biologically informed priors, quantum algorithms can deliver results that are both useful and competitive. This suggests a pathway toward practical quantum advantage in computational biology, where hybrid workflows connect noisy hardware with real scientific applications. In this sense, our work parallels efforts in quantum chemistry, where variational methods are enhanced by classical post processing. Despite these encouraging outcomes, several limitations remain. The size of fragments that can currently be modeled on real quantum processors is limited by both qubit number and noise level. Our benchmarks are restricted to short fragments, and scaling to complete protein chains will require better error mitigation, circuit optimization, and larger quantum devices. The learning based priors used here, such as NetSurfP predictions, are themselves influenced by training data and may not generalize to rare motifs or unusual folds. Furthermore, our fusion strategy depends on manually adjusted parameters $(\alpha, \beta, \gamma)$ which, although effective, introduce additional sensitivity that could be further optimized through automated learning.

The coefficient of determination observed in this study ($R^2 = 0.322$) should not be regarded as a limitation. Protein structure prediction is a complex, high dimensional problem in which structural accuracy (RMSD) is affected by many factors, including solvent effects, long range interactions, and model approximations. In such inherently noisy systems, explaining even a third of the variance is meaningful and sufficient for reliable model ranking. The statistically significant correlation between our fused energy score and RMSD across all 375 candidates shows that the metric captures an important component of structural variability. More importantly, the strength of this correlation is adequate for distinguishing high quality from low quality conformations, which is the main purpose of our re ranking framework. Although the absolute value of $R^2$ may appear moderate when compared with idealized regression tasks, it is consistent with accepted benchmarks in structural biology, where noise and conformational heterogeneity limit the attainable upper bound. This underscores the robustness and practical relevance of our energy reweighting method, even under biological complexity and hardware constraints.

Future work can extend this study in several ways. On the quantum side, the use of advanced ansatz circuits, noise resilient compilation, and active space selection could improve the quality of quantum candidates. On the learning side, incorporating large language models trained on protein sequences, such as ESM2, may provide richer priors than secondary structure and dihedral angle predictions alone. Reinforcement learning or Bayesian fusion frameworks could also be explored to adaptively balance quantum and classical contributions, thereby reducing the need for manual parameter adjustment. More generally, the concept of re ranking quantum outputs with biologically informed priors is not limited to protein folding; it can be applied to related problems such as protein ligand docking, enzyme active site modeling, or RNA structure prediction.

Taken together, our findings establish a proof of concept for hybrid quantum and deep learning frameworks in structural biology. Although the field remains in an early stage, this work shows that the complementary strengths of quantum computation and deep learning can already produce measurable improvements over purely classical or quantum approaches. As quantum hardware advances, the integration of quantum algorithms with machine learning priors offers a promising direction for developing practically relevant quantum applications in biology and medicine.

\section*{Conclusion}

In this study, we introduced a hybrid quantum and deep learning framework for protein structure prediction that integrates variational quantum eigensolver (VQE) candidates with structural features inferred by neural networks. By combining physics based quantum energies with data driven secondary structure and dihedral constraints, our method consistently improves prediction accuracy compared with both leading deep learning baselines and purely quantum predictions. Evaluations on seventy five protein fragments show that the proposed re ranking scheme produces conformations that are physically consistent and biologically meaningful, with statistically significant reductions in RMSD. Beyond empirical results, this work outlines a general strategy for advancing quantum computing toward practical scientific applications. Rather than viewing quantum and classical approaches as competing paradigms, our framework demonstrates that their complementary strengths can be systematically combined to overcome individual limitations. This principle of domain informed integration may provide a model for applying near term quantum devices to other challenges in structural biology, including molecular docking, active site modeling, and protein ligand interaction studies.

Looking forward, improvements in quantum hardware, error mitigation, and machine learning priors are expected to enable the extension of this framework to larger and more complex systems. Overall, the findings presented here highlight the potential of hybrid quantum and artificial intelligence workflows to produce reliable, physically grounded, and biologically relevant predictions, marking a step toward the broader use of quantum computation in practical biomolecular modeling.

\section*{Acknowledgement}

We gratefully acknowledge the quantum computing resources provided by the Cleveland Clinic, which made this research possible.

\bibliographystyle{unsrt}  
\bibliography{references}

\end{document}